\documentclass[usenatbib,letterpaper,hyperref]{mn2e}
\usepackage[labelfont=bf,labelsep=period]{caption}
\usepackage[totalwidth=480pt, totalheight=680pt]{geometry}
\usepackage{times}
\usepackage{epsfig}
\usepackage{amsmath}
\usepackage{amssymb}
\usepackage{url}
\usepackage{aas_macros}
\def\eqsim{\mathrel{\raise0.35ex\hbox{$\scriptstyle =$}\kern-0.6em
    \lower0.40ex\hbox{{$\scriptstyle \sim$}}}}
\def\gtrsim{\mathrel{\raise0.35ex\hbox{$\scriptstyle >$}\kern-0.6em
    \lower0.40ex\hbox{{$\scriptstyle \sim$}}}}
\def\lesssim{\mathrel{\raise0.35ex\hbox{$\scriptstyle <$}\kern-0.6em
    \lower0.40ex\hbox{{$\scriptstyle \sim$}}}}
\def\HI{H\,{\sc i}}
\def\Ha{H\,$\alpha$}
\def\barolo{$^{3{\rm D}}${\sc barolo}}
\def\apostle{APOSTLE}
\def\eagle{EAGLE}
\def\things{THINGS}
\def\littlethings{LITTLE~THINGS}

\title[Non-circular motions and rotation curve diversity]{Non-circular motions and the diversity of dwarf galaxy rotation curves}
\author[K. A. Oman et al.]{
  \newauthor Kyle A. Oman$^{1,2}$\thanks{koman@astro.rug.nl}, Antonino Marasco$^{2,3}$, Julio F. Navarro$^{1,4}$, Carlos S. Frenk$^{5}$,
  \newauthor Joop Schaye$^{6}$, Alejandro Ben{\'i}tez-Llambay$^{5}$
  \\
$^{1}$ Department of Physics \& Astronomy, University of Victoria, Victoria, BC, V8P 5C2, Canada\\
$^{2}$ Kapteyn Astronomical Institute, University of Groningen, Postbus 800, NL-9700 AV Groningen, The Netherlands\\
$^{3}$ ASTRON, Netherlands Institute for Radio Astronomy, Postbus 2, 7900 AA, Dwingeloo, The Netherlands\\
$^{4}$ Senior CIfAR Fellow\\
$^{5}$ Institute for Computational Cosmology, Department of Physics, University of Durham, South Road, Durham DH1 3LE, United Kingdom\\
$^{6}$ Leiden Observatory, Leiden University, PO Box 9513, NL-2300 RA Leiden, the Netherlands
}
\date{\today}

\begin{document}
\label{firstpage}
\maketitle

\begin{abstract} %235/250 word limit
 We use mock interferometric \HI\ measurements and a conventional tilted-ring modelling procedure to estimate circular velocity curves of dwarf galaxy discs from the \apostle\ suite of $\Lambda$CDM cosmological hydrodynamical simulations. The modelling yields a large diversity of rotation curves for an \emph{individual galaxy} at fixed inclination, depending on the line-of-sight orientation. The diversity is driven by non-circular motions in the gas; in particular, by strong bisymmetric fluctuations in the azimuthal velocities that the tilted-ring model is ill-suited to account for and that are difficult to detect in model residuals. Large misestimates of the circular velocity arise when the kinematic major axis coincides with the extrema of the fluctuation pattern, in some cases mimicking the presence of kiloparsec-scale density `cores', when none are actually present. The thickness of \apostle\ discs compounds this effect: more slowly-rotating extra-planar gas systematically reduces the average line-of-sight speeds. The recovered rotation curves thus tend to underestimate the true circular velocity of \apostle\ galaxies in the inner regions. Non-circular motions provide an appealing explanation for the large apparent cores observed in galaxies such as DDO~47 and DDO~87, where the model residuals suggest that such motions might have affected estimates of the inner circular velocities. Although residuals from tilted ring models in the simulations appear larger than in observed galaxies, our results suggest that non-circular motions should be carefully taken into account when considering the evidence for dark matter cores in individual galaxies.
\end{abstract}

\section{Introduction}
\label{SecIntro}

The `cusp-core problem' is a long-standing controversy that arises when contrasting the steep central density profiles of cold dark matter haloes (`cusps') predicted by N-body simulations \citep{1996ApJ...462..563N,1997ApJ...490..493N} with the mass profiles inferred from disc galaxy rotation curves after subtracting the contribution of the stellar and gaseous (baryonic) components \citep[][and see \citealp{2010AdAst2010E...5D} for a review]{1994ApJ...427L...1F, 1994Natur.370..629M}. The comparison usually involves fitting a power-law density profile to the dark matter contribution in the innermost resolved region of the rotation curve. The power-law slope is then compared with that of simulated CDM haloes at similar distances from the centre. 

Although legitimate in principle, this procedure is in practice fraught with difficulties.  One difficulty is that rotation speeds rapidly approach zero near the centre, which implies that inferences about the dark matter cusp are made by fitting small rotation velocities at small radii, a regime where even small errors can have a large influence on the results.  A further difficulty is that, in order to recover the dark mass profile, one must remove the baryonic contribution to the circular velocity, which involves making assumptions about relatively poorly known parameters, such as, for example, the mass-to-light ratio of the stars, and the $X_{\rm CO}$ parameter used to infer the molecular hydrogen distribution from CO observations. However, even with extreme assumptions for the values of these parameters, the recovered innermost slopes are in general shallower than predicted for CDM haloes \citep[e.g.][]{1997MNRAS.290..533D,1999PhDT........27S,2011AJ....141..193O}.

The reason for this emphasis on the innermost slope may be traced to early N-body simulation work \citep{1991ApJ...378..496D,1996ApJ...462..563N,1997ApJ...490..493N}, which reported that the the central cusp of CDM halo density profiles was at least as steep as $\rho \propto r^{-1}$. Any slope measured to be shallower than that could then argued to be in conflict with CDM, an idea that has guided many studies of inner rotation curves since.

Our understanding of this issue, however, has evolved, largely as a result of improved simulations and of better constraints on the cosmological parameters. Indeed, observations of the cosmic microwave background and of large-scale galaxy clustering have now constrained the cosmological parameters to great precision \citep{2016A&A...594A..13P}. At the same time, cosmological  simulations have improved to the point that the halo scaling parameters, their dependence on mass, as well as their scatter, are now well understood \citep[e.g.][]{2014MNRAS.441..378L}. 

Because of these advances, the dark matter contribution to a rotation curve can now be predicted {\it at essentially all relevant radii}  once a single parameter is specified for a halo, such as the maximum circular velocity, $V_{\rm max}$. Since baryons can dissipate and add mass to the inner regions, this dark matter contribution may be regarded as a minimum mass (or, equivalently, a minimum circular velocity) at each  radius. The advantage of focussing on this minimum velocity is that it allows the theoretical predictions to be directly confronted with observations at radii not too close to the centre, where rotation curves are less prone to uncertainty and where the theoretical predictions are less vulnerable to artefact.

This is the approach we adopted in an earlier study \citep{2015MNRAS.452.3650O}, where we proposed to reformulate the cusp-core problem as an `inner mass deficit' problem that afflicts galaxies where the inner circular velocities fall below the minimum expected from the dark matter alone. At a fiducial radius\footnote{We note that there is nothing exceptional about this choice; it is just a compromise radius that both simulations and observations resolve well for a wide range of galaxy masses.} of $2\,{\rm kpc}$, a number of galaxies have rotation speeds much lower than predicted, given their $V_{\rm max}$. As discussed in that paper, this mass deficit {\it does not affect all galaxies} but it does affect galaxies of all masses, from dwarfs to massive discs, and varies widely from galaxy to galaxy at fixed $V_{\rm max}$. 

In some cases, the deficit at $2\,{\rm kpc}$ is so pronounced that it far exceeds the total baryonic mass of the system.  This argues (at least in those extreme examples) against the idea that the deficit  is caused by dark matter `cores' produced by the baryonic assembly of the galaxy \citep[see, e.g.,][see also the review of \citealp{2014Natur.506..171P}]{1996MNRAS.283L..72N,2005MNRAS.356..107R,2008Sci...319..174M,2012MNRAS.422.1231G,2012MNRAS.421.3464P,2014ApJ...786...87B,2015MNRAS.454.2092O,2016MNRAS.456.3542T}. Indeed, these baryon-induced modifications are in general modest, and the effect restricted to a small range of galaxy masses \citep{2014MNRAS.437..415D,2015MNRAS.454.2981C}. In addition, we note that baryon-induced cores are {\it not} a general prediction, but rather a result of {\it some} implementations of star formation and feedback in galaxy formation simulations. Indeed, simulations like those from the \eagle\ \citep{2015MNRAS.446..521S}, \apostle\ \citep{2016MNRAS.457.1931S} and Illustris \citep{2014MNRAS.444.1518V} projects are able to reproduce most properties of the galaxy population without producing any such cores.

An alternative is that gas rotation curves do not faithfully trace the circular velocity in the inner regions of some galaxies and have therefore been erroneously interpreted as evidence for `cores'. This possibility has been repeatedly raised in the past: early observations were subject to concerns around beam smearing in the case of \HI\ data \citep[][and references therein]{2009A&A...493..871S}, and centering, alignment, and seeing in the case of \Ha\ slit spectroscopy \citep{2003ApJ...583..732S,2005AJ....129.2119S}. Such worries have largely been laid to rest with the advent of new, high-resolution datasets that often combine \Ha, \HI, and CO maps to yield 2D gas velocity fields of excellent angular resolution \citep[e.g.,][]{2006ApJS..165..461K,2008AJ....136.2563W,2012AJ....144..134H,2014ApJ...789...63A}.

Other concerns, however, remain. Observed velocity data must be processed through a model before they can be contrasted with theoretical predictions, and a number of modelling issues have yet to be properly understood. The main purpose of modelling the data is to infer the speed of a hypothetical circular orbit, $V_{\rm circ}(r)$, which can then be directly compared with the mass distribution interior to radius $r$ predicted by the simulations. Observations, however, can at best only constrain the average azimuthal speed of the gas at each radius, usually referred to as the `rotation speed', $V_{\rm rot}(r)$. In general, $V_{\rm rot}\neq V_{\rm circ}$, and corrections must be applied. 

One such correction concerns the support provided by pressure gradients and velocity dispersion of the gas. This depends on the gas surface density profile, as well as on the gas velocity dispersion and its radial gradient, in a manner akin to the familiar `asymmetric drift' that affects the average rotation speed of stars in a disc \citep[e.g.][]{2007ApJ...657..773V}. Although the corrections are approximate, for galaxies with $V_{\rm max}\gtrsim 30\,{\rm km}\,{\rm s}^{-1}$ the changes they imply are usually too small to compromise the results \citep[e.g.][]{2011AJ....141..193O}.

Non-circular motions in the gas, on the other hand, are a greater concern. Although these are likely ubiquitous at some level, they are seldom considered in the modelling. One reason for this is that there is no simple and general way of assessing the effect of non-circular motions, which may affect both the estimates of $V_{\rm rot}$ {\it and} the translation of $V_{\rm rot}$ into $V_{\rm circ}$.  As a result, $V_{\rm rot}(r)$ is often used as a direct measure of $V_{\rm circ}(r)$ without further correction. Although this practice may sometimes be acceptable, such as in the case of the 19 galaxy discs from the \things\ survey studied in detail by \citet{2008AJ....136.2720T}, it can also lead to erroneous conclusions (as we shall see below) and must be carefully scrutinized for each individual galaxy.

The case of NGC~2976 provides a sobering example. Obvious asymmetries in the velocity field led \citet{2003ApJ...596..957S} to use a harmonic decomposition of the velocity field, where circular gas `rings' were allowed to have non-zero radial velocity (i.e., they may be expanding or contracting), in addition to the usual rotation speed. With this assumption, a `tilted-ring' model \citep{1974ApJ...193..309R} can reproduce the observed gas velocity field quite accurately, yielding a well-defined mean azimuthal velocity as a function of radius, $V_{\rm rot}(r)$. (And, of course, non-zero radial velocities too.)

A visual inspection of the velocity field of NGC~2976 shows that it  differs significantly in contiguous quadrants, but is roughly antisymmetric in diagonally opposite ones. This is a clear signature of eccentric, rather than circular, gas orbits \citep{2006MNRAS.373.1117H}, which led \citet{2007ApJ...664..204S} to model this galaxy assuming the presence of a radially-coherent bisymmetric ($m=2$) velocity pattern, as expected in a barred galaxy, or when the halo potential is triaxial. This model also fits the data quite well, but yields a rather different radial dependence for $V_{\rm rot}(r)$, especially near the centre. 

In addition to this model degeneracy, {\it in neither case} does the derived `rotation curve' $V_{\rm rot}(r)$ -- the mean azimuthal rotation of the gas -- trace the circular velocity, $V_{\rm circ}(r)$. Translating $V_{\rm rot}$ into $V_{\rm circ}$ in such cases requires a dynamical gas flow model of the whole galaxy, something that can only be accomplished when specific assumptions are made about the ellipticity of the gravitational potential, its dependence on radius, and/or the radial dependence of its phase angle. Given these complexities, it is not surprising that non-circular motions are usually not modelled in detail and instead treated as a source of error.

We explore these issues here by analysing synthetic \HI\ observations of $33$ simulated central galaxies with $60 < V_{\rm max}/{\rm km}\,{\rm s}^{-1} < 120$ selected from the \apostle\ suite of cosmological hydrodynamical simulations \citep{2016MNRAS.457..844F,2016MNRAS.457.1931S}. We use \barolo, a tilted-ring processing tool that has recently been used to derive the rotation curves of galaxies in the \littlethings\ survey \citep{2017MNRAS.466.4159I}. We begin in Sec.~\ref{SecPWork} with a brief review of earlier work on kinematic models of galactic discs. In Sec.~\ref{SecSims} we briefly describe the simulations and how we construct our synthetic observations. We compare these with measurements of real galaxies in Sec.~\ref{SubsecKinProps} to show that the two are similar in terms of several kinematic and symmetry metrics. In Sec.~\ref{SecModels} we describe the process of fitting kinematic models to our synthetic observations. In Sec.~\ref{SecResults} we present our main results: in Sec.~\ref{SecOrientation} we demonstrate that the recovered rotation curve depends sensitively on the orientation of non-circular motions in the disc with respect to the line of sight, and in Sec.~\ref{SecDiscThickProj} we show that the mixing of different annuli and layers in the disc along the line of sight is a further source of error. In Sec.~\ref{SecIMD} we show that these non-circular motions can lead to `inner mass deficits' comparable to those reported by \citet{2015MNRAS.452.3650O}, and compare with observed galaxies in Sec.~\ref{SecIMDObs}. Finally, we discuss the implications of our findings and summarize our conclusions in Sec.~\ref{SecConc}.

\section{Prior work}
\label{SecPWork}

We are hardly the first to suggest that non-circular motions can substantially impact the rotation curve modelling process. The pioneering work of \citet{1991wdir.conf...40T} was followed up by \citet{1994ApJ...436..642F} and \citet{1997MNRAS.292..349S} to extract the signature, in projection, of perturbations to circular orbits. This work showed that, when expanded into harmonic modes, each mode of order $m$ gives rise to patterns of order $m\pm1$ in projection. 

Projection effects thus introduce degeneracies in the interpretation of non-circular motions, which add to others introduced by errors in geometric parameters such as the systemic velocity, centroid, inclination, and position angle of the kinematic principal axes. Some of the degeneracies may be lifted if the amplitude of the perturbations is small, which allows the epicyclic approximation to be used to constrain the various amplitudes and phases, but this approach is of little help when perturbations are a substantial fraction of the mean azimuthal velocity, as is often the case near the galaxy centre.

\citet{2004ApJ...617.1059R} discuss the kinematic modelling of a simulated barred galaxy (their `Model I'). In this case, the bar is strong enough to drive down the mean rotation velocity substantially. The authors find that the rotation curve they measure for the system depends sensitively on the orientation of the bar, with the apparent rotation falling far below the circular velocity of the system when the bar is aligned along the major axis of the galaxy in projection, just the case when asymmetries in the projected velocity field are hardest to detect. They also note that even very small systematic errors in the velocity, on the order of $10$~per~cent, are enough to cause large changes in the inferred dark matter density profile slope.

A similar cautionary tale is told by \citet{2007ApJ...657..773V}, who analyse synthetic observations of simulations of isolated galaxies set up initially in equilibrium. They argue that non-circular motions, coupled with the extra support provided by gas pressure, are enough to explain the slowly rising rotation curves of NGC~3109 and NGC~6822, two galaxies where the inferred dark matter profiles depart substantially from that expected from $\Lambda$CDM. Their conclusion is that cuspy profiles are actually consistent with the data once these effects are taken into account.

\citet{2007ApJ...664..204S} extended earlier work to account for perturbations that are large compared with the mean azimuthal velocity, a regime where the epicyclic approximation fails. Their model applies to bisymmetric ($m=2$) perturbations of radially-constant phase, as in the case of a bar or a triaxial dark halo. Applied to a real galaxy, the method enables estimates of the mean orbital speed from a velocity map, even when strongly non-axisymmetric. These authors also caution that this is just a measure of the average azimuthal speed around a circle, and not a precise indicator of centrifugal balance.  In other words, $V_{\rm rot}\neq V_{\rm circ}$ when non-circular motions are not negligible. In this case, the only way to estimate $V_{\rm circ}(r)$ is to find a non-axisymmetric model that produces a fluid dynamical flow pattern matching the observed one.

Given these complexities, it is an illuminating exercise to `observe' simulated galaxies and analyse them on an even footing with analogous observed data. This is the approach we adopt in this paper. Previous comparisons between synthetic rotation curves from gas dynamic simulations and real data include the recent work by \citet{2016MNRAS.462.3628R}, who analysed simulations of $4$ idealized galaxies which have dark matter cores, and by \citet{2017MNRAS.466...63P}, who analysed a set of $6$ simulated isolated galaxies which retain their initial dark matter cusps. We draw a sample of galaxies from the \apostle\ suite of cosmological hydrodynamical simulations \citep{2016MNRAS.457.1931S,2016MNRAS.457..844F}, construct synthetic `observations' of their \HI\ gas kinematics, and apply the same `tilted-ring' modelling commonly adopted to analyse the kinematics of nearby discs. None of the \apostle\ galaxies have `cores', which simplifies the interpretation: any possible deviations between the recovered $V_{\rm rot}(r)$ and the known $V_{\rm circ}(r)$ are either due to the fact that the gas is not truly in centrifugal equilibrium, or to inadequacies in the modelling of the data.

\section{Simulations}
\label{SecSims}

\subsection{The \apostle\ simulations}
\label{SubsecApostle}

The \apostle\footnote{A Project Of Simulations of The Local Environment.} simulation suite comprises $12$ volumes selected from a large cosmological volume and resimulated using the zoom-in technique \citep{1996ApJ...472..460F,2003MNRAS.338...14P,2013MNRAS.434.2094J} with the full hydrodynamics and galaxy formation treatment of the `Ref' model of the \eagle\ project \citep{2015MNRAS.446..521S,2015MNRAS.450.1937C}. The regions are selected to resemble the Local Group of galaxies in terms of the mass, separation and kinematics of two haloes analogous to the Milky Way and M~31, and relative isolation from more massive systems. Full details of the simulation setup and target selection are available in \citet{2016MNRAS.457..844F} and \citet{2016MNRAS.457.1931S}; we summarize a few key points here.

\eagle, and by extension \apostle, use the pressure-entropy formulation of smoothed particle hydrodynamics \citep{2013MNRAS.428.2840H} and the numerical methods from the {\sc anarchy} module [Dalla Vecchia et al. (in preparation); see \citealp{2015MNRAS.446..521S} for a short summary]. The galaxy formation model includes subgrid recipes for radiative cooling \citep{2009MNRAS.393...99W}, star formation \citep{2004ApJ...609..667S,2008MNRAS.383.1210S}, stellar and chemical enrichment \citep{2009MNRAS.399..574W}, energetic stellar feedback \citep{2012MNRAS.426..140D}, and cosmic reionization \citep{2001cghr.confE..64H,2009MNRAS.399..574W}. The stellar feedback is calibrated to reproduce approximately the galaxy stellar mass function and the sizes of $M_\star>10^8\,{\rm M}_\odot$ galaxies at $z=0$. \eagle\ also includes black holes and AGN feedback. In \apostle\ these processes are neglected, which is reasonable for the mass scale of interest here \citep{2015MNRAS.450.1937C,2017MNRAS.465...32B}.

The \apostle\ volumes are simulated at three resolution levels, labelled \mbox{AP-L3} (similar to the fiducial resolution of the \eagle\ suite), \mbox{AP-L2} (similar to the `high resolution' \eagle\ runs) and an even higher resolution level, \mbox{AP-L1}. Each resolution level represents an increase by a factor of $\sim 12$ in mass and $\sim 2$ in force softening over the next lowest level. Typical values of particle mass, gravitational softening, and other numerical parameters vary slightly from volume to volume; representative values are shown in Table~\ref{tab_APparams}. All $12$ volumes have been simulated at \mbox{AP-L2} and \mbox{AP-L3} resolution, but only $5$ volumes: V1, V4, V6, V10 \& V11 have thus far been simulated at \mbox{AP-L1}. \apostle\ assumes the WMAP-7 cosmological parameters \citep{2011ApJS..192...18K}: $\Omega_{\rm m} = 0.2727$, $\Omega_\Lambda=0.728$, $\Omega_{\rm b}=0.04557$, $h=0.702$, $\sigma_8=0.807$.

The {\sc subfind} algorithm \citep{2001MNRAS.328..726S,2009MNRAS.399..497D} is used to identify structures and galaxies in the \apostle\ volumes. Particles are first grouped into friend-of-friends (FoF) haloes by iteratively linking particles separated by at most $0.2\times$ the mean interparticle separation \citep{1985ApJ...292..371D}; gas and star particles are attached to the same FoF halo as their nearest dark matter particle. Saddle points in the density distribution are then used to separate substructures, and particles which are not gravitationally bound to substructures are removed. The end result is a collection of groups each containing at least one `subhalo'; the most massive subhalo in each group is referred to as `central'; others are `satellites'. In this analysis we focus exclusively on central objects as satellites are subject to additional dynamical processes which complicate their interpretation.

We label our simulated galaxies according to the resolution level, volume number, FoF group and subgroup, so for instance \mbox{AP-L1-V1-8-0} corresponds to resolution \mbox{AP-L1}, volume V1, FoF group 8 and subgroup 0 (the central object). We focus primarily on the \mbox{AP-L1} resolution. At this resolution level the circular velocity curves of our galaxies of interest (defined below) are numerically converged at all radii $\gtrsim 700\,{\rm pc}$, as defined by the criterion of \citet[][for further details pertaining to the numerical convergence of the \apostle\ simulations see \citealp{2015MNRAS.452.3650O,2016MNRAS.457.1931S,2017MNRAS.469.2335C}]{2003MNRAS.338...14P}.

\begin{table}
  \caption{Summary of the key parameters of the \apostle\ simulations used in this work. Particle masses vary by up to a factor of $2$ between volumes at a fixed resolution `level'; the median values below are indicative only \citep[see][for full details]{2016MNRAS.457..844F}. Details of the WMAP-7 cosmological parameters used in the simulations are available in \citet{2011ApJS..192...18K}.\label{tab_APparams}}
  \begin{tabular}{llll}
    \hline
    & \multicolumn{2}{l}{Particle masses (${\rm M}_\odot$)} & Max softening \\
    Simulation & DM & Gas & length (${\rm pc}$)\\
    \hline
    AP-L3 & $7.3\times10^6$ & $1.5\times10^6$ & $711$ \\
    AP-L2 & $5.8\times10^5$ & $1.2\times10^5$ & $307$ \\
    AP-L1 & $3.6\times10^4$ & $7.4\times10^3$ & $134$ \\
    \hline
  \end{tabular}
\end{table}

\subsection{Galaxy sample selection}
\label{SubsecSample}

We select $33$ galaxies from the \apostle\ simulations for further consideration based on two criteria. First, as noted above, we restrict ourselves to the highest \mbox{AP-L1} resolution level so that the central regions of the galaxies, which are of particular interest in the context of the cusp-core problem, are sufficiently well resolved. Second, we choose galaxies in the interval $60 < V_{\rm max}/{\rm km}\,{\rm s}^{-1} < 120$, where $V_{\rm max}={\rm max}(V_{\rm circ}(R))$. The lower bound ensures that the gas distribution of the galaxies is well-sampled ($\gtrsim 10^4$ gas particles contribute to the \HI\ distribution of each galaxy). The upper bound is chosen to exclude massive galaxies were the baryonic component dominates the kinematics. Indeed, the kinematics of all galaxies in our sample are largely dictated by their dark matter haloes.

\begin{figure*}
  {\leavevmode \includegraphics[width=2.12\columnwidth]{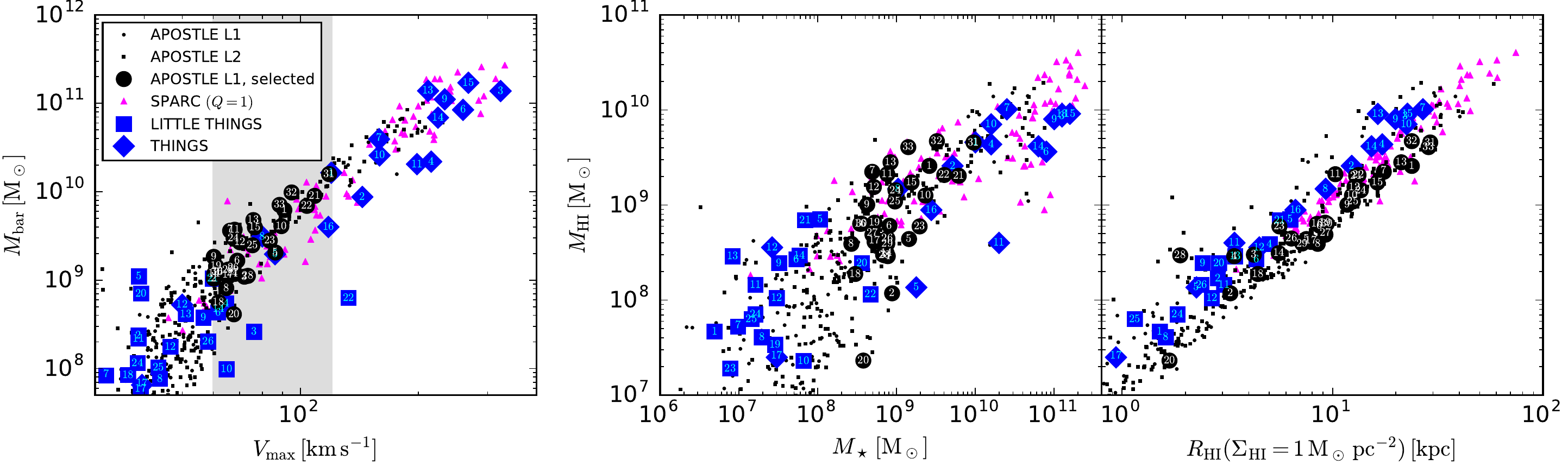}}
  \caption{\emph{Left: }Baryonic Tully-Fisher relation (BTFR) for \apostle\ galaxies at resolution \mbox{AP-L1} (black circles) and \mbox{AP-L2} (black squares). For comparison we also show the BTFR for the SPARC sample of galaxies (magenta triangles) and the \things\ (blue squares, numbering corresponds to Table~\ref{tab_thingsprops}) and \littlethings\ (blue diamonds, see also Table~\ref{tab_thingsprops}) galaxies. In all cases, we assume $M_{\rm gas} = 1.4M_{\rm HI}$. All \mbox{AP-L1} galaxies in the range $60<V_{\rm max}/{\rm km}\,{\rm s}^{-1}<120$ (indicated by the grey shaded band) are selected for further analysis and shown with larger, numbered symbols (see Table~\ref{tab_APprops}). \emph{Centre: }\HI\ mass -- stellar mass relation; symbols and numbering are as in the left panel. \emph{Right: }\HI\ mass--size relation. Sizes are defined as the radius where the \HI\ surface density drops to $1\,{\rm M}_\odot\,{\rm pc}^{-2}$ ($\approx 10^{20}\,{\rm atoms}\,{\rm cm}^{-2}$). Symbols and numbering are as in the left panel. \label{fig_size_mass}}
\end{figure*}

\apostle\ galaxies have realistic masses, sizes, and velocities. This is shown in Fig.~\ref{fig_size_mass}, where we compare the simulated sample (large black points) with observational data from the Spitzer Photometry and Accurate Rotation Curves (SPARC) database \citep{2016AJ....152..157L} and the \things\ \citep{2008AJ....136.2563W} and \littlethings\ \citep{2012AJ....144..134H} surveys. For the SPARC galaxies, we plot only those with the highest quality flag ($Q=1$). For the \things\ and \littlethings\ surveys, we plot only those galaxies which were selected for kinematic modelling by the survey teams (\things: \citealp{2008AJ....136.2648D,2011AJ....141..193O}; \littlethings: \citealp{2015AJ....149..180O}).

\apostle\ galaxies comfortably match three key scaling relations. The left panel of Fig.~\ref{fig_size_mass}  shows the baryonic Tully-Fisher relation (BTFR). The quantity plotted on the horizontal axis varies by dataset: for \apostle\ galaxies we show the maximum of the circular velocity curve, $V_{\rm circ}(r)=\sqrt{GM(<r)/r}$, for SPARC galaxies we show the asymptotically flat rotation velocity, whereas  for \things\ \& \littlethings\ galaxies we show the maximum of the rotation curve. The baryonic masses are, in all cases, calculated as $M_{\rm bar}=M_\star + 1.4 M_{\rm HI}$ \citep[e.g.][and see Sec.~\ref{SubsecSimCubes} for the method used to calculate the \HI\ masses]{2012AJ....143...40M}. Our selection in $V_{\rm max}$ is highlighted by the shaded vertical band. It is clear from this panel that the BTFR of \apostle\ galaxies is in good agreement with the observed scaling, provided that the observed velocities trace the maximum circular velocity of the halo \citep[see; e.g.,][for a more in-depth discussion of this point]{2017MNRAS.464.2419S,2016MNRAS.460.3610O}.

The middle panel shows the \HI\ mass -- stellar mass relation. The simulated galaxies once again lie comfortably within the scatter of the observed relation. The right panel shows the \HI\ mass -- size relation, where the size is defined as the radius at which the \HI\ surface density, $\Sigma_{\rm HI}$, drops below $1\,{\rm M}_\odot\,{\rm pc}^{-2}\,(\approx 10^{20}\,{\rm atoms}\,{\rm cm}^{-2})$. \apostle\ galaxies seem to have, at fixed \HI\ mass, slightly larger sizes ($\sim 0.2\,{\rm dex}$) than observed. The offset in size shown in the right-hand panel of Fig.~\ref{fig_size_mass} should be of little consequence to our analysis.

\subsection{Synthetic \HI\ data cubes}
\label{SubsecSimCubes}

For each simulated galaxy in our sample we carry out a synthetic \HI\ observation, as follows. First, we compute an \HI\ mass fraction for each gas particle in the central galaxy, following the prescription of \citet{2013MNRAS.430.2427R} for self-shielding from the metagalactic ionizing background radiation, and including an empirical pressure-dependent correction for the molecular gas fraction, as detailed in \citet{2006ApJ...650..933B}. 

Second, we adopt a  coordinate system centred on the potential minimum of the galaxy, and choose a $z$-axis aligned with the direction of $\vec{L}_{\rm HI}$, the specific angular momentum vector of the \HI\ gas disc. The velocity coordinate frame is chosen such that the average (linear) momentum of the \HI\ gas in the central $500\,{\rm pc}$ is zero. A viewing angle inclined by $60^\circ$ relative to the $z$-axis is adopted, with random azimuthal orientation. Each galaxy is placed in the Hubble flow at a nominal distance of $3.7\,{\rm Mpc}$, the median distance of galaxies in the \littlethings\ sample \citep{2012AJ....144..134H}. We choose an arbitrary position on the sky at $(0^{\rm h}\,0^{\rm m}\,0.0^{\rm s}, +10^\circ\,0^\prime\,0.0^{\prime\prime})$ and adopt an `observing setup' similar to that used in the \littlethings\ survey, with a $6\,{\rm arcsec}$ circular Gaussian beam and $1024^2$ pixels spaced $3\,{\rm arcsec}$ apart. This yields an effective physical resolution (FWHM) of $\sim 110\,{\rm pc}$. We use a velocity channel spacing of $4\,{\rm km}\,{\rm s}^{-1}$ and enough channels to accommodate comfortably all of the galactic \HI\ emission.

The gas particles are spatially smoothed with the $C^2$ \citet{Wendland1995} smoothing kernel used in the \eagle\ model. The integral of the kernel over each pixel is approximated by the value at the pixel centre. Provided the pixel size is $\leq \frac{1}{2}$ the smoothing length, this approximation is accurate to better than 1~per~cent; we explicitly verify that this condition is satisfied. We also verified that omitting this smoothing step does not significantly change our main results.

In the velocity direction, the $21\,{\rm cm}$ emission is modelled with a Gaussian line profile centred at the particle velocity and a fixed width of $7\,{\rm km}\,{\rm s}^{-1}$, which models the (unresolved) thermal broadening of the \HI\ line \citep[e.g.][]{2017MNRAS.466...63P}. Our main results are insensitive to the precise width we choose for the line, provided it is $\lesssim 12\,{\rm km}\,{\rm s}^{-1}$, because then the integrated \HI\ profile is dominated by the dispersion in the particle velocities. Each particle contributes flux proportionally to its \HI\ mass, i.e. the gas is assumed to be optically thin. Finally, the synthetic data cube is convolved along the spatial axes with the `beam', implemented as a $6\,{\rm arcsec}$ circular Gaussian kernel. The completed cube is saved in the {\sc fits} format \citep{2010A&A...524A..42P} with appropriate header information.

\begin{figure*}
  {\leavevmode \includegraphics[width=2\columnwidth]{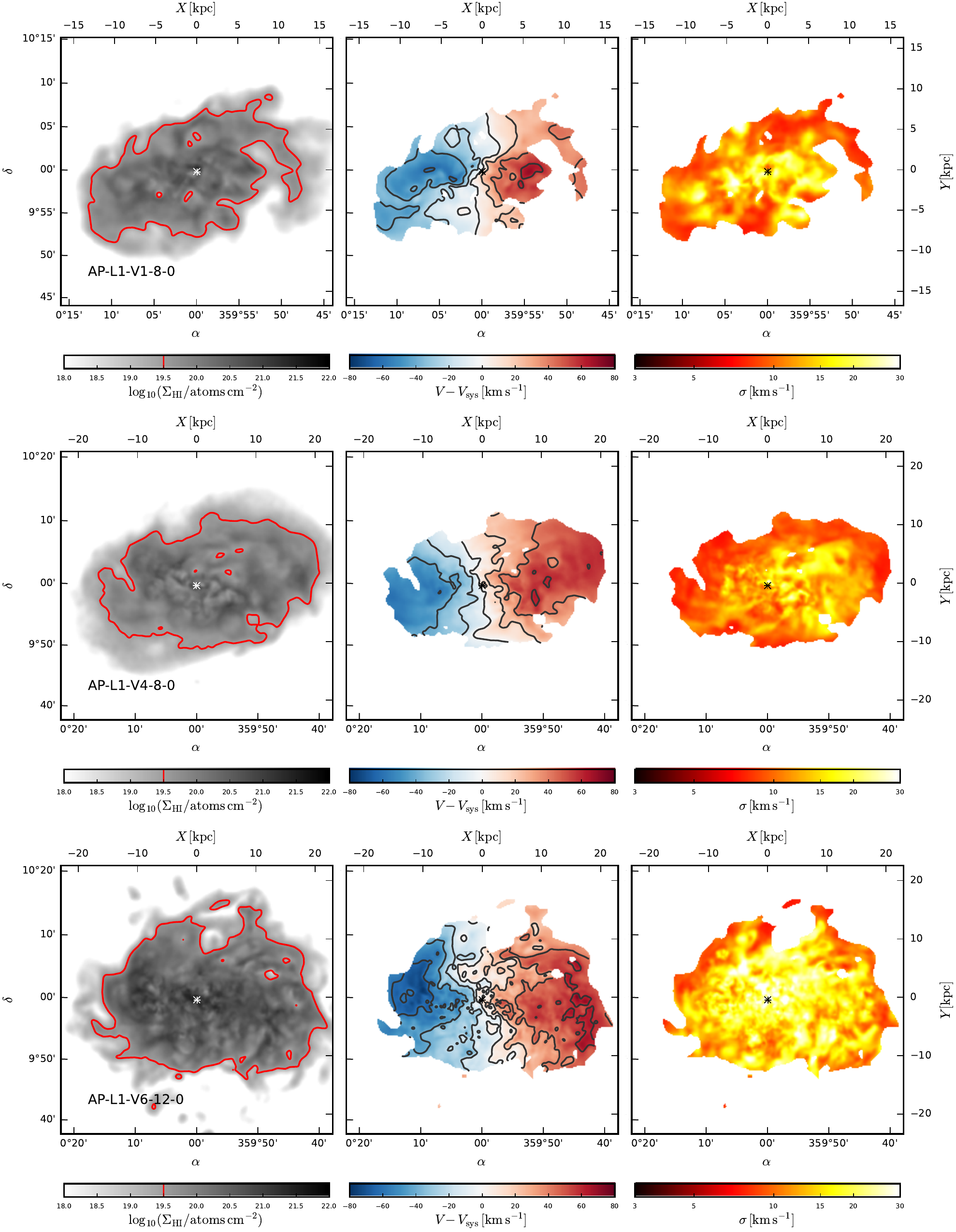}}
  \caption{From left to right: $0^{\rm th}$ moment (surface density), $1^{\rm st}$ moment (flux-weighted mean velocity) and $2^{\rm nd}$ moment (flux-weighted velocity dispersion) maps for three objects in our sample of \apostle\ galaxies. The galaxies are placed at an arbitrary sky position at a distance of $3.7\,{\rm Mpc}$, inclination of $60^\circ$ and position angle of $270^\circ$ (angle East of North to the approaching side), where the angular momentum vector of the \HI\ disc is taken as the reference direction. The $1^{\rm st}$ and $2^{\rm nd}$ moment maps are masked to show only pixels where the surface density exceeds $10^{19.5}\,{\rm atoms}\,{\rm cm}^{-2}$ (indicated by the red line in the surface density map). Contours on the $1^{\rm st}$ moment map correspond to the tick locations on the colour bar. The `$\times$' marks the location of the potential minimum, which is well-traced by the peak of the stellar distribution, marked `$+$'. See also Appendix~\ref{AppSupp}.\label{fig_mom_maps}}
\end{figure*}

In Fig.~\ref{fig_mom_maps} we illustrate the synthetic observations of three of our simulated galaxies. The left column shows the surface density ($0^{\rm th}$~moment) maps. The red contour marks the $\log_{10}(\Sigma_{\rm HI}/{\rm atoms}\,{\rm cm}^{-2})=19.5$ isodensity contour. This is about $0.5\,{\rm dex}$ deeper than the typical limiting depth of observations in the \things\ and \littlethings\ surveys of $\sim 10^{20}\,{\rm atoms}\,{\rm cm}^{-2}$. We choose this because  galaxies in our sample are slightly larger than observed ones, by roughly $\sim 0.2\,{\rm dex}$ in $M_{\rm HI}$--$R_{\rm HI}$. In light of this, a slightly deeper nominal limiting column density allows for more reasonable comparisons than a strict cut at $10^{20}\,{\rm atoms}\,{\rm cm}^{-2}$. The central column shows the line-of-sight velocity ($1^{\rm st}$~moment) maps\footnote{We show intensity weighted mean (IWM) velocity fields. The choice of velocity field type can have a significant impact on the fit rotation curve for techniques that model the velocity field directly \citep{2008AJ....136.2648D}. For our purposes, however, the choice of velocity field impacts only the visualization of the data because our model of choice, \barolo, models the full data cube.}, and the right column the velocity dispersion ($2^{\rm nd}$~moment) maps.

\subsection{Kinematics properties of simulated and observed galaxies}
\label{SubsecKinProps}

\begin{figure*}
  {\leavevmode \includegraphics[width=2.1\columnwidth]{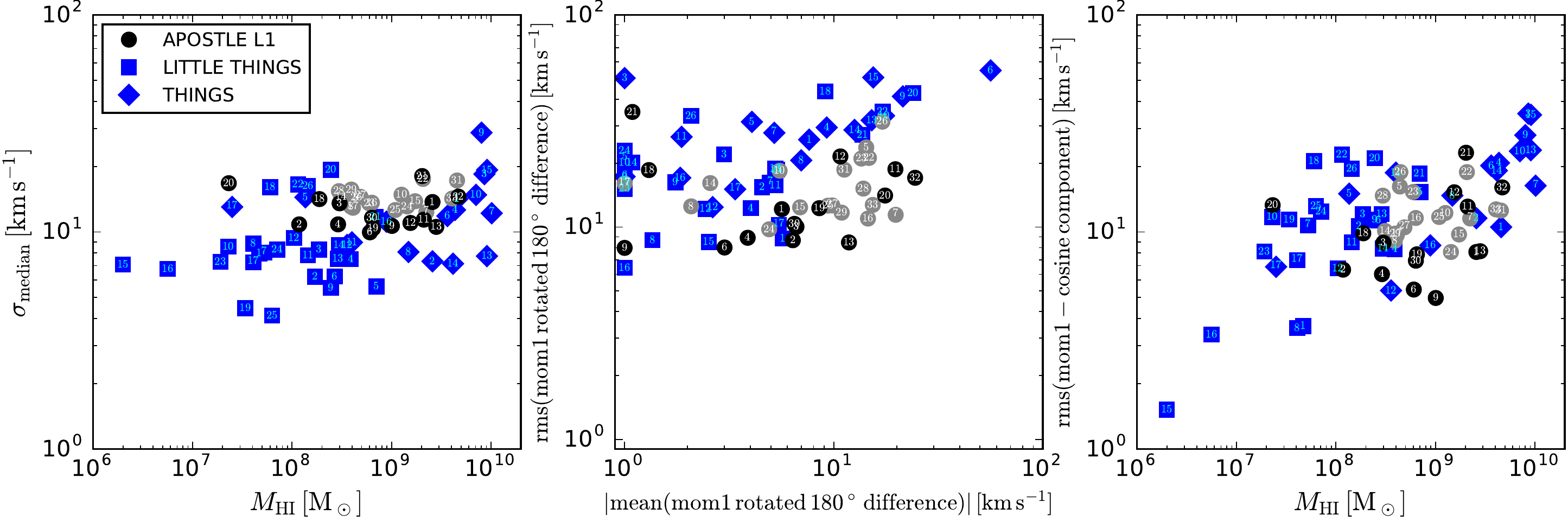}}
  \caption{Diagnostics comparing the kinematics of observed and simulated galaxies. In all panels, numbering is as in Fig.~\ref{fig_size_mass}. \emph{Left: }Median velocity dispersion as measured along the line of sight as a function of \HI\ mass. For the \apostle\ galaxies, the median is calculated across all pixels with $\log_{10}(\Sigma_{\rm HI}/{\rm atoms}\,{\rm cm}^{-2}) > 19.5$; for the \things\ and \littlethings\ galaxies it is computed across all pixels in the S/N masked second moment map. Light grey symbols correspond to galaxies which we flag as kinematically disturbed (see Fig.~\ref{fig_profiles} and Sec.~\ref{SubSecGasRotVel}). \emph{Centre: }As a measure of the symmetry of the velocity field, the first moment (mean velocity field) of each data cube is rotated $180^\circ$ about its centre and subtracted from itself (with a sign change); here we plot the rms against the absolute mean offset from $0$ of the pixels. Pixels which overlap a pixel with no velocity measurement after rotation are discarded. See Fig.~\ref{fig_symmetry_ill} and Appendix~\ref{SecApp2} for an illustration and further explanation of this measurement. \emph{Right: }The rms about zero of the residual velocity field, derived by subtracting a simple axisymmetric model (Eq.~\ref{EqSimpleKinModel}) from the original velocity field, as a function of \HI\ mass. See Fig.~\ref{fig_symmetry_ill_cosine} and Appendix~\ref{SecApp2} for an illustration and further explanation of this measurement.\label{fig_symmetry}}
\end{figure*}

Are the kinematic properties of simulated galaxies broadly consistent with observed ones? We have already seen in Fig.~\ref{fig_size_mass} that \apostle\ galaxies have structural parameters that follow scaling laws similar to observed discs, but it is important to check that they also resemble observations in their internal kinematics. We explore this using three simple metrics that we can apply both to the publicly available moment maps\footnote{We use the `robust weighted', not the `natural weighted', maps \citep{2008AJ....136.2648D}, though both give very similar results.} of observed galaxies as well as to our simulated data cubes with minimal extra processing. The observational maps are provided cleaned of noise, with low signal-to-noise pixels masked out. We approximate this by masking in the simulated maps all pixels where  the \HI\ column density drops below $10^{19.5}\,{\rm atoms}\,{\rm cm}^{-2}$ (see Fig.~\ref{fig_mom_maps}).

The first metric is the median velocity dispersion along the line of sight (i.e. the median of all unmasked pixels in the $2^{\rm nd}$~moment map) as a function of \HI\ mass, which we show in the left panel of Fig.~\ref{fig_symmetry}. \apostle\ galaxies are shown by grey/black symbols, whereas galaxies from the \things\ and \littlethings\ surveys are shown with blue squares and diamonds (note that we plot only those galaxies regular enough to have been selected for mass modelling by the survey teams). Each symbol has a number that identifies the galaxy as listed in Tables~\ref{tab_APprops} and \ref{tab_thingsprops}. At given \HI\ mass the simulated galaxies have slightly larger velocity dispersions than observed galaxies, but the difference is less than a factor of two on average.

The second metric estimates the symmetry of the $1^{\rm st}$~moment maps (i.e., the line-of-sight velocity field). This is computed by rotating a map by $180^\circ$ about the galaxy centre and subtracting it from the unrotated fields (with a change of sign so that in the perfectly symmetric case the residual would be zero everywhere). The mean of the residual map indicates whether there is an offset in the average velocity of the approaching and receding sides of the galaxy; its rms is a crude estimate of the lack of circular symmetry of the velocity field. As shown by the middle panel of  Fig.~\ref{fig_symmetry}, \apostle\ and observed galaxies seem to deviate from perfect axisymmetry by similar amounts.

Finally, the third metric uses a measure of the residuals produced by subtracting a very simple kinematic model from the $1^{\rm st}$~moment map. Assuming a single inclination, position angle and systemic velocity for each galaxy (as listed in Table~\ref{tab_thingsprops}), we fit the function:
\begin{equation}
V_{\rm LoS}(\phi) = V_{\rm sys} + V_0\cos(\phi-\phi_0)\label{EqSimpleKinModel}
\end{equation}
to a series of concentric, inclined `rings' (ellipses in projection). We use the same ring spacings as \citet{2008AJ....136.2648D,2011AJ....141..193O,2015AJ....149..180O}, typically about $130\,{\rm pc}$. $V_0$ and $\phi_0$ are free\footnote{The freedom in $\phi_0$ means that, strictly speaking, we are not removing a pure rotation field. We recall that the purpose of this measurement is to compare synthetic and real data cubes, and the measurement is made identically in both cases.} parameters fit to each ring independently. The residual map is then analysed as for the preceding metric: its rms is shown as a function of \HI\ mass in the right-hand panel of Fig.~\ref{fig_symmetry}. As in the other cases, the simulated and observed galaxies are nearly indistinguishable according to these metrics. 

These results, together with those shown earlier in Fig.~\ref{fig_size_mass}, give us confidence that the kinematics of the simulated galaxies are, to zeroth order, similar enough to those of their observed counterparts to warrant applying similar analysis tools.

\begin{figure*}
  {\leavevmode \includegraphics[width=2\columnwidth]{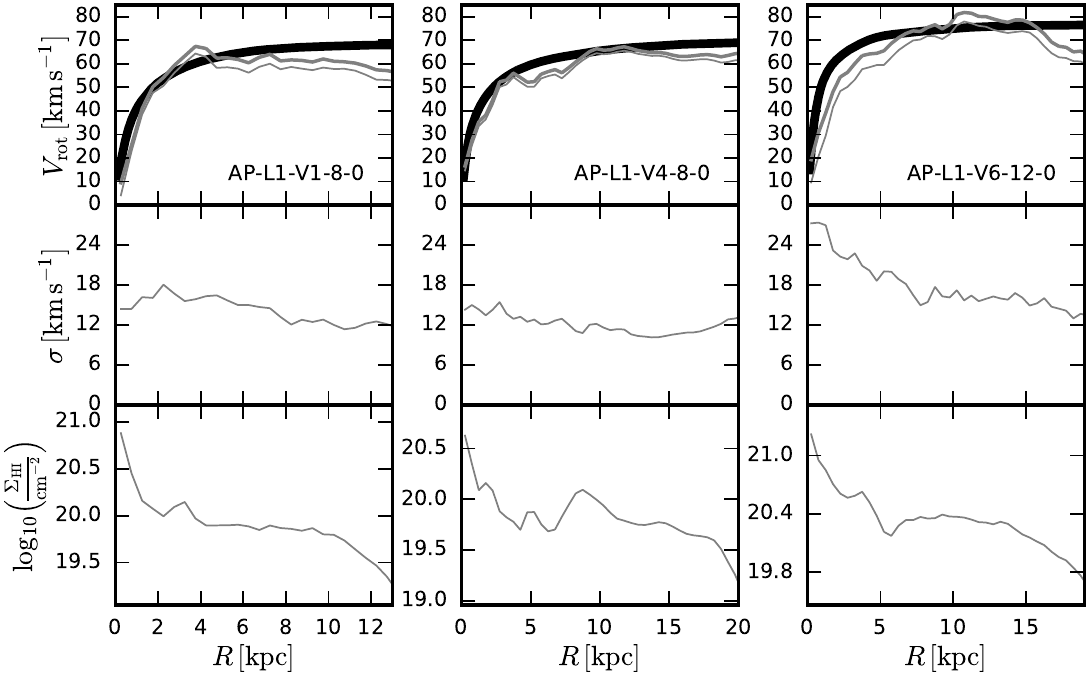}}
  \caption{\emph{First row: }Circular velocity curves (heavy black lines) and mean azimuthal velocity of \HI\ gas (thin grey lines) for three of the simulated galaxies in our sample. The gas rotation velocity corrected for pressure support (see Sec.~\ref{SubsecADC}) is shown with the thick grey line. Since we have chosen our sample to have relatively large $V_{\rm max}>60\,{\rm km}\,{\rm s}^{-1}$, such corrections are typically quite small. We flag galaxies in which the pressure-corrected velocity at $2\,{\rm kpc}$ differs from the circular velocity by more than $15$~per~cent, such as \mbox{AP-L1-V6-12-0}, as kinematically disturbed. \emph{Second row: }\HI\ velocity dispersion profiles for the same galaxies, including both the thermal (subparticle) and interparticle contributions to the velocity dispersion, and calculated assuming isotropy as $1/\sqrt{3}$ of the 3D velocity dispersion at each radius. \emph{Third row: }\HI\ surface density profiles for the same galaxies. The plot is truncated at the radius enclosing $90$~per~cent of the \HI\ mass, which is typically very close to the radius where the surface density drops below our nominal limiting $\Sigma_{\rm HI}$ depth of $10^{19.5}\,{\rm atoms}\,{\rm cm}^{-2}$.\label{fig_profiles}}
\end{figure*}

\section{Kinematic modelling}
\label{SecModels}

\subsection{Tilted-ring model}
\label{SecTiltedRing}

The standard tool for kinematic modelling of disc galaxy velocity fields is known as a `tilted-ring' model \citep{1974ApJ...193..309R}. In such a model, a disc is represented as a series of rings of increasing size. The properties of each ring are described by a set of parameters which can be categorized as geometric (radius, width, thickness, centroid, inclination, position angle, systemic velocity) and physical (surface density, rotation velocity, velocity dispersion). A number of publicly-available tilted-ring models exist; we use here the \barolo\footnote{\url{http://editeodoro.github.io/Bbarolo/}, we used the latest version available at the time of writing: 1.3 (github commit d54e901).} software package \citep[for a detailed description see][]{2015MNRAS.451.3021D}. 

Whereas most older versions of tilted-ring models only use the first few moments of the kinematics -- the surface density and velocity fields, and in some cases the velocity dispersion field -- \barolo\ belongs to a class of more recent tools that model the full data cube directly, and therefore nominally utilize all available kinematic information. The software has many configurable parameters; we discuss our choices for several of the most important ones below, and in Table~\ref{tab_baroloparams} we summarize the full configuration used.

\subsection{Parameter choices}
\label{SubsecParamChoices}

The most important parameters of the model are those that define the handling of the geometric parameters of each ring. When applied to projections of \apostle\ galaxies, and in order to facilitate convergence, we provide \barolo\ with a `correct' guess of $i=60^\circ$ for the inclination angle (and allow it to deviate by no more than $15^\circ$ from this value). We also initialize the software with the `correct' guess for the position angle of the rings ($270^\circ$ counter-clockwise from North), and allow deviations of no more than $20^\circ$. Providing reasonably accurate initial guesses (within $\sim 15^\circ$) for these two parameters is, unfortunately, necessary for the fitting procedure to converge to a correct solution \citep{2015MNRAS.451.3021D}. For real galaxies, these must be estimated from the geometry of either the gas or stellar distribution. The inclination and position angles that would be derived from the gas isodensity contours for our sample of \apostle\ galaxies typically differ from the `true' values by less than the maximum variations we allow in the fitting routine. The ring widths are fixed at $14.1\,{\rm arcsec}$, corresponding to a physical separation of $250\,{\rm pc}$ at the distance of $3.7\,{\rm Mpc}$ chosen for our synthetic observations.  

We fix the centre of each ring to the density peak of the projected stellar distribution of the galaxy. This coincides, within a few pixels ($<3\,{\rm px} \sim 10\,{\rm arcsec}$), with the minimum potential centre returned by the {\sc subfind} algorithm \citep{2001MNRAS.328..726S,2009MNRAS.399..497D}. For simplicity, the systemic velocity is fixed at $257\,{\rm km}\,{\rm s}^{-1}$, determined from the distance as $V_{\rm sys}=H_0D$. The initial guesses for the rotation speed and velocity dispersion of each ring are set to $30$ and $8\,{\rm km}\,{\rm s}^{-1}$, respectively. These initial guesses have little impact on the final fits to the rotation curve and velocity dispersion profile.

We fix the thickness of the rings at $2\,{\rm arcsec} = 40\,{\rm pc}$. This is much thinner than the actual thicknesses of the simulated gas discs, where the half-mass height can reach $\sim 1\,{\rm kpc}$.  Modelling thick discs is a well-known limitation of tilted ring models. Future codes may be able to capture better the vertical structure of discs \citep[e.g.][]{2017MNRAS.466.4159I}, but for the present we are bound by the limitations of current implementations.

We model each galaxy out to the radius enclosing 90~per~cent of its \HI\ mass. This roughly coincides with the $\log_{10}(\Sigma_{\rm HI}/{\rm atoms}\,{\rm cm}^{-2})=19.5$ isodensity contour, and is, in all cases, extended enough to reach the asymptotically flat (maximum) portion of the circular velocity curve. 

\subsection{Fitting procedure}

Using the parameter choices outlined above (see also Table~\ref{tab_baroloparams}), the tilted ring model is fit to each galaxy in two stages \citep[e.g.][]{2017MNRAS.466.4159I}. In the first stage the free parameters are the rotation speed, velocity dispersion, inclination and position angle of each ring (in \barolo's `locally normalized' mode the surface brightness is not explicitly fit). The inclination and position angle profiles are then smoothed with a low-order polynomial fit and, in a second stage, the rotation speeds and velocity dispersions of the rings are fit again with the geometric parameters held fixed at their smoothed values.

\subsection{Correction for pressure support}
\label{SubsecADC}

The procedure above yields the mean azimuthal velocity of the galaxy as a function of radius, $V_{\rm rot}(r)$. This is usually smaller than the true circular velocity because the gas may be partially supported by `pressure' forces. We therefore correct the rotation speeds as in, e.g., \citet{2007ApJ...657..773V}:
\begin{equation}
  V_{\rm circ}^2 = V_{\rm rot}^2 - \sigma^2\frac{{\rm d}\log(\Sigma_{\rm HI}\sigma^2)}{{\rm d\log R}}
\label{EqAsymmDrift}
\end{equation}
where $\Sigma_{\rm HI}$ is the surface density of the \HI\ gas and $\sigma$ is the component of the velocity dispersion along the line of sight. This formulation of the pressure support correction is the one most commonly employed in the rotation curve literature. It is often called the `asymmetric drift' correction because its formulation is analogous to the familiar correction that applies to (collisionless) stellar discs, although the two corrections have different physical origins \citep[see, e.g.,][for a discussion]{2017MNRAS.466...63P}. This correction is not, strictly speaking, correct, as it assumes a single gas phase and that no bulk flows are present in the disc. Neither assumption holds exactly, of course, but this formula is enough to assess whether pressure forces make an important contribution to the disc kinematics.

We measure the surface density along the (projection of) each of the best-fitting rings directly from the synthetic data cubes. In practice, we measure the gradient of  the `pressure' profile $\Sigma_{\rm HI}\sigma^2$ using the following fitting function ($\alpha$, $(\Sigma_{\rm HI}\sigma^2)_0$ and $R_0$ are free parameters):
\begin{equation}
  \frac{\Sigma_{\rm HI}\sigma^2}{(\Sigma_{\rm HI}\sigma^2)_0} = \frac{(R_0 + 1)}{R_0 + e^{\alpha R}}
\end{equation}
This is the same functional form used in recent analyses of the \things\ and \littlethings\ galaxies\footnote{\Citet{2008AJ....136.2648D} make no mention of pressure support corrections in their analysis of \things\ galaxies, though for the majority of the galaxies in their sample the correction would be expected to be very small.} \citep{2011AJ....141..193O,2015AJ....149..180O,2017MNRAS.466.4159I}. 

\section{Results}
\label{SecResults}

\begin{figure*}
  {\leavevmode \includegraphics[width=2\columnwidth]{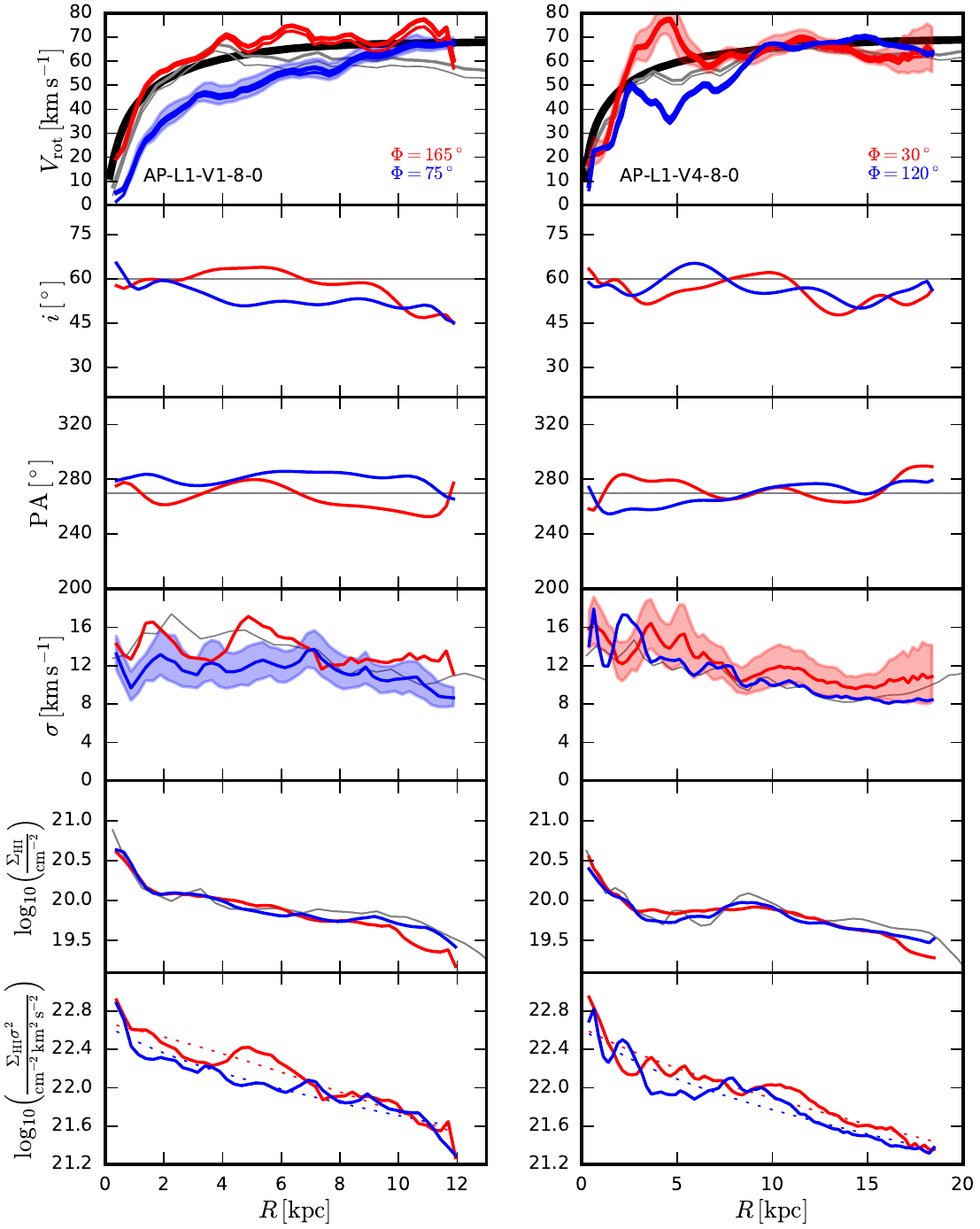}}
  \caption{Kinematic modelling of two of the galaxies shown in Fig.~\ref{fig_profiles} (left and centre columns). Fits for two orientations of each galaxy, labelled by $\Phi$ (see Fig.~\ref{fig_az_dep}), are shown by the red and blue curves, offset from each other by a $90^\circ$ rotation about the galactic pole. \emph{First row: }Rotation curves: circular velocity curve (thick black), gas azimuthal velocity (thin grey), same corrected for pressure support (thick grey), kinematic model with regularized geometric parameters (thin coloured), same corrected for pressure support (thick coloured) with errors estimated by \barolo\ (shaded area -- for clarity only shown for one orientation). \emph{Second row: }Inclination profiles: nominal inclination (thin grey), regularized inclination profile (coloured). \emph{Third row: }As second row, but for the position angle profile. \emph{Fourth row: }Velocity dispersion profiles; velocity dispersion calculated directly from simulation particle distribution (grey; Eq.~\ref{EqDispBreakdown}), kinematic model with regularized geometric parameters (coloured) with errors (shaded area). \emph{Fifth row: }\HI\ surface density profiles; surface density calculated directly from simulation particle distribution (grey), surface density along the projection of each ring defined by the regularized inclination and position angle profiles (coloured). \emph{Sixth row: } $\Sigma_{\rm HI}\sigma^2$ profiles; the profiles shown with coloured lines in the fourth and fifth rows are combined and fit with a simple function (dotted lines, see Sec.~\ref{SubsecADC}) for use in calculating the pressure support correction for the (thin solid coloured) rotation curves shown in the first row. See also Appendix~\ref{AppSupp}.\label{fig_barolo_fits}}
\end{figure*}

\subsection{Gas rotation velocities}
\label{SubSecGasRotVel}

Before discussing the application of the tilted-ring model described in the previous section to \apostle\ galaxies, we begin by comparing the mean azimuthal speed of the gas in the disc plane, $V_{\rm rot}(r)$, with the true circular velocity of the system, $V_{\rm circ}(r)$. The purpose of this exercise is to weed out cases where the gas is patently out of equilibrium, since our main goal is to examine the possible shortcomings of the tilted-ring model for galaxies where the disc is close to equilibrium. This is, very roughly, analogous to the common practice of omitting galaxies with obvious kinematic irregularities (mergers, strong bars or tidal features) from rotation curve observing campaigns or analyses. We stress that the `equilibrium' criterion used here is indicative only, and cannot be replicated in observed galaxies, where the true circular velocity profile is unknown. The distinction between equilibrium and non-equilibrium galaxies is only adopted in order to simplify the interpretation of our analysis, and not to compare with observations. In particular, we note that several of the galaxies which we discard as out-of-equilibrium would very likely be included in observational samples of relaxed galaxies.

The rotation profile was measured using the \HI\ mass-weighted mean azimuthal velocity of gas particles in a series of $2\,{\rm kpc}$ thick, $500\,{\rm pc}$ wide cylindrical shells aligned along the disc plane. The velocity dispersion profile was measured using the same series of rings. The 1D line-of-sight gas velocity dispersion, $\sigma$, results from the contribution from the (isotropic) thermal pressure plus that of the `bulk'  motion of the gas; i.e.
\begin{equation}
  \sigma = \sqrt{\frac{k_{\rm B}T}{\mu m_{\rm p}} + \frac{1}{3}\left(\sigma_\phi^2 + \sigma_r^2 + \sigma_z^2\right)} \label{EqDispBreakdown}
\end{equation}
where $k_{\rm B}$ is Boltzmann's constant, $T$ is the particle temperature, $\mu$ is its mean molecular weight, $m_{\rm p}$ is the proton mass, and $\sigma_\phi$, $\sigma_R$ and $\sigma_z$ are the azimuthal, radial and vertical components of the gas particle velocity dispersion. Both components are reflected in the synthetic data cubes (Sec.~\ref{SubsecSimCubes}), though in practice the `bulk' component always dominates by a factor $>2$.

We show three examples from \apostle\ in Fig.~\ref{fig_profiles}, where each column refers to a different galaxy, and, from top to bottom, each panel shows, respectively, the rotation speed, the 1D velocity dispersion, and the \HI\ surface density profiles. The thick black curve in the top panels denotes $V_{\rm circ}(r)$; the thin grey curve $V_{\rm rot}(r)$; and the thick grey curve the `pressure-corrected' rotation speed, as in Eq.~\ref{EqAsymmDrift}. Note that, as anticipated in Sec.~\ref{SecIntro}, the pressure corrections are usually small.

If we focus on the inner (rising) part of the rotation curves shown in Fig.~\ref{fig_profiles}, we see that the mean gas rotation speed closely traces the circular velocity in two of the three galaxies. The gas rotation curve of the galaxy in the rightmost column, on the other hand, deviates quite strongly from $V_{\rm circ}(r)$ in the inner regions, indicating that this galaxy has likely undergone a recent perturbation that has pushed the gas component temporarily out of equilibrium. Galaxies like the latter are highlighted in Fig.~\ref{fig_symmetry} by a lighter shade of grey and are excluded from the analysis that follows, leaving $15$ galaxies. (The actual criterion adopted is that the pressure-corrected $V_{\rm rot}$ differs from $V_{\rm circ}$ by more than $15$~per~cent at a fiducial radius of $2\,{\rm kpc}$.)

\subsection{Orientation and tilted-ring rotation curves}
\label{SecOrientation}

Having excluded galaxies where the inner gas disc is clearly out of equilibrium, we proceed to model the remaining galaxies using \barolo. Although the inclination is fixed at $60^{\circ}$ in all synthetic observations, there is still freedom to choose a second angle to define the line of sight. Fig.~\ref{fig_barolo_fits} shows two \barolo\ fits to the `equilibrium' galaxies in Fig.~\ref{fig_profiles}, as obtained for two different line-of-sight orientations.  These were not chosen at random, but have instead been selected to demonstrate the importance of orientation effects on the rotation curves of seemingly `equilibrium' galaxies in \apostle.

The tilted-ring modelling returns rotation curves that at times underestimate significantly the mean azimuthal speed of the gas (see blue curves), and, consequently, its circular velocity. The situation changes when the galaxy is rotated by $90^\circ$, keeping the same inclination: in this case  (shown in red) the inferred rotation speeds are substantially higher, and at times even exceed the true circular velocity of the system. Note that the difference between the red and blue rotation curves is much greater than the `errors' that the model assigns to the recovered $V_{\rm rot}(r)$; these are shown by the shaded area\footnote{Errors shown are as estimated by \barolo: the model parameters are resampled around the best-fitting values to determine the variations required to change the model residual by 5~per~cent. This yields an error similar to what might be derived from differences between the approaching and receding sides of the galaxy \citep{2015MNRAS.451.3021D}.} around the rotation curves. (Shaded areas are only shown for a couple of curves for clarity.)

The differences in the recovered rotation curves cannot be ascribed to variations in the inclination (a difference of $10^\circ$ at $i=60^\circ$ only changes $V_{\rm rot}$ by $\sim10$~per~cent), or in the velocity dispersion (differences of $\lesssim 4\,{\rm km}\,{\rm s}^{-1}$), or in the pressure correction inferred by the model, as may be seen from the other panels in Fig.~\ref{fig_barolo_fits}. The model actually recovers these parameters quite well, which implies that the orientation dependence must be due to the presence of large-scale, coherent non-circular motions in the plane of the disc.

\subsection{Non-circular motions and orientation effects}
\label{subsec_ncm_orientation}

\begin{figure*}
  {\leavevmode \includegraphics[width=2\columnwidth]{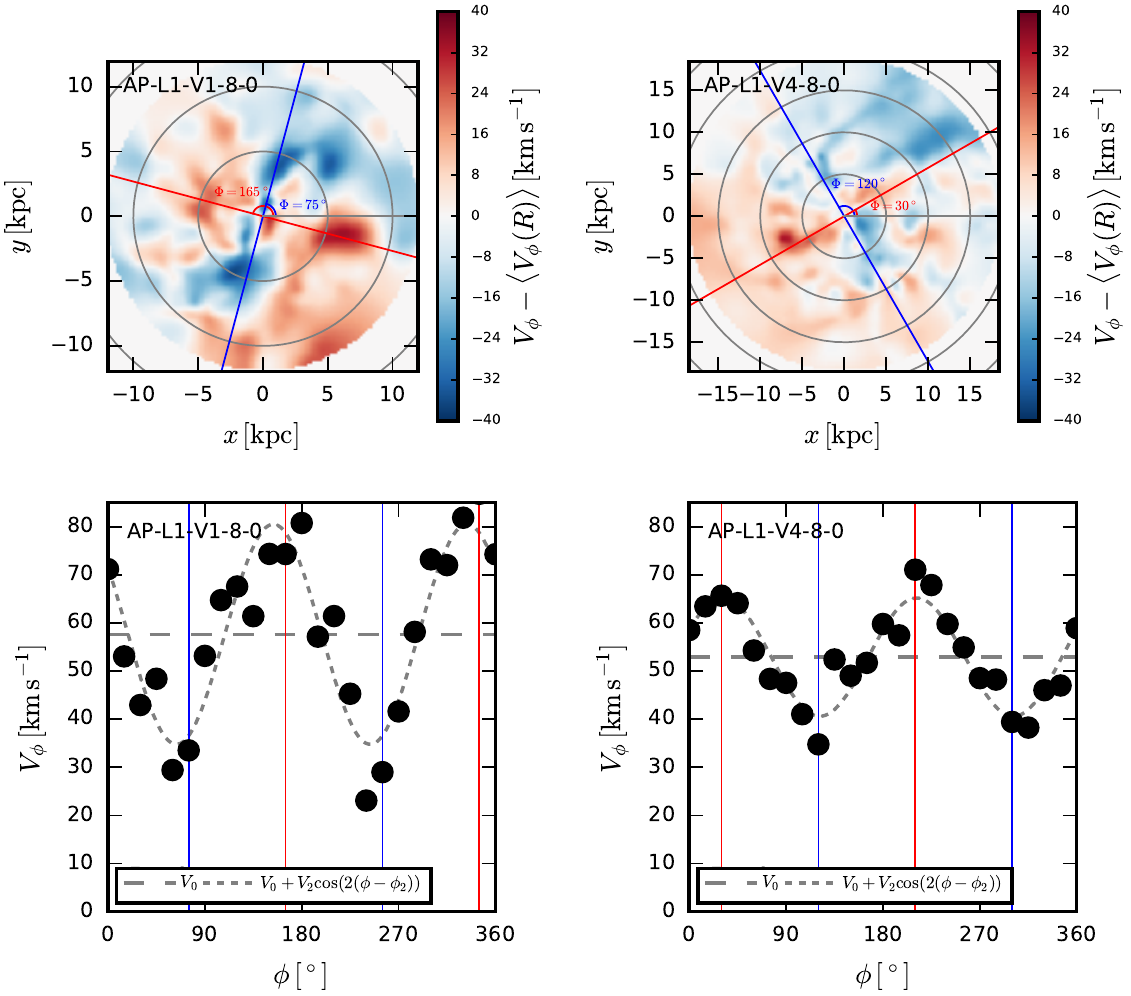}}
  \caption{\emph{First row: }Face-on maps of the residual azimuthal motions (after subtracting the mean rotation as a function of radius) in the disc plane for the two galaxies shown in Fig.~\ref{fig_barolo_fits}. The red and blue lines correspond to the directions that lie along the major axis of the projections modelled and shown with the lines of corresponding colour in Fig.~\ref{fig_barolo_fits}. We label the projection orientation $\Phi$ according to its angular offset from the $x$-axis, as illustrated. The grey circles are drawn at intervals of $5\,{\rm kpc}$. \emph{Second row: }Azimuthal velocity at $5\,{\rm kpc}$ as a function of azimuth (black points). The best-fitting $m=0$ and $2$ terms of a Fourier series are shown with broken line styles. The vertical coloured lines correspond to the directions along the lines of the same colours in the upper panels, and coincide approximately with the peaks and troughs of the $m=2$ mode. This alignment, though imperfect, extends to larger and smaller radii as well.\label{fig_az_dep}}
\end{figure*}

The presence of large-scale non-circular motions is illustrated in Fig.~\ref{fig_az_dep}, where the top panels show, for each galaxy, the residual azimuthal motions in the discs after subtracting the mean rotation at each radius, where the mean is measured directly from the simulation particle information. The line of nodes (i.e., the projected major axis) of the two projection axes shown in Fig.~\ref{fig_barolo_fits} are illustrated by the lines, with corresponding colours. Note the presence of a clear radially coherent bisymmetric pattern in the residual velocities, which explains the results obtained by \barolo\ in projection. The bisymmetric perturbation (which resembles that of a slowly-rotating bar-like pattern) is caused by the triaxial nature of the dark matter halo that hosts the galaxy \citep{2004MNRAS.355..794H}, as we discuss in detail in a companion paper \citep{2018MNRAS.476.2168M}.

When the projected major axis slices through the minima of the pattern (blue lines) the recovered rotation velocities underestimate the true rotation speed; the opposite happens when the major axis slices through the two maxima of the residual map (red lines). This is because, in projection, most of the information about the rotation velocity is contained in sight lines near the major axis -- gas rotating faster or slower than the average on the projected major axis drives the rotation curve up or down, respectively.

The bottom panels of Fig.~\ref{fig_az_dep} further illustrate the non-circular motion pattern. The points indicate the rotation speed as a function of azimuthal angle at a radius $R=5\,{\rm kpc}$ (innermost grey ring in the upper panels). We fit the $m=0$ and $m=2$ terms of a Fourier series:
\begin{equation}
V(\phi)=\sum_m V_m\cos[m(\phi - \phi_m))] \label{EqFourier}
\end{equation}
with amplitudes $V_m$ and phases $\phi_m$, to these points, and plot the two terms separately with dashed line styles. In both cases there is a strong $m=2$ component. The maxima of this mode align with the projection axis drawn in red in the upper panels (red vertical lines in lower panels); the minima align with the direction drawn in blue.

In principle there may also be an $m=1$ component in the non-circular motions; however, at any given radius its amplitude is degenerate with the assumed systemic velocity. Here we have adopted a systemic velocity that minimizes the $m=1$ term in the harmonic expansion at $5\,{\rm kpc}$ in order to focus on the bisymmetric component.

\begin{figure*}
  {\leavevmode \includegraphics[width=2\columnwidth]{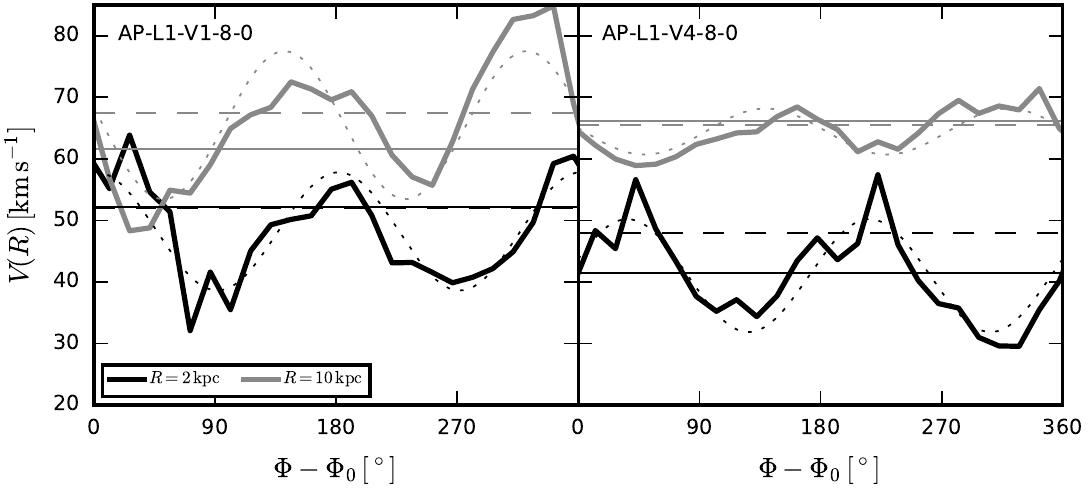}}
  \caption{Rotation velocity at $2$ and $10\,{\rm kpc}$ as recovered by \barolo\ as a function of projection axis $\Phi$, including pressure support corrections. The reference direction $\Phi_0$ is defined as the nominal direction of the maximum of the $m=2$ pattern in the upper quadrants, i.e. the red line in the upper panels of Fig.~\ref{fig_az_dep}. $\Phi_0=165^\circ$ and $30^\circ$ for \mbox{AP-L1-V1-8-0} and \mbox{AP-L1-V4-8-0}, respectively. The horizontal dashed lines show the circular velocity at the same radii, while the horizontal solid lines shows the mean rotation velocity of the \HI\ gas, also at the same radii, measured directly from the simulation particles, and corrected for pressure support. We expect the fit rotation speed to vary proportionally to $\cos(2(\Phi-\Phi_0))$; we show the best-fitting $V_0+V_2\cos(2((\Phi-\Phi_0)-\Phi'))$ with a dotted line (note the additional freedom $\Phi'$ in the phase). In general $\Phi'\neq 0$ because in some cases the $m=2$ pattern (Fig.~\ref{fig_az_dep} upper panels) is, at some radii, not exactly aligned along the direction defined by $\Phi_0$.\label{fig_az_rot}}
\end{figure*}

Fig.~\ref{fig_az_rot} confirms unambiguously the effect of this $m=2$ pattern on the rotation curve recovered by \barolo. Here we show the rotation speed recovered by the tilted-ring model at two different radii ($R=2$ and $10\,{\rm kpc}$) as a function of the orientation of the line of sight (keeping the inclination always fixed at $i=60^{\circ}$).  Clearly, as the orientation varies the inferred rotation speed varies as expected from a dominant $m=2$ mode (i.e., two maxima and two minima as the galaxy is spun by $360^{\circ}$). Note that the phase of the modulation varies between the two radii, slightly in the case of \mbox{AP-L1-V1-8-0} and more strongly for \mbox{AP-L1-V4-8-0}. This indicates that the phase of the $m=2$ mode shifts gradually with radius, as may be corroborated by visual inspection of the residual maps in Fig.~\ref{fig_az_dep}.

\subsection{Non-circular motions in projection}
\label{SubsecObsApplicability}

Harmonic modulations of the velocity field (of order $m$) are not always easily discernible in projection, where they are mapped into a combination of $m \pm 1$ modes. The analogue in projection along the line of sight of Eq.~\ref{EqFourier} is:
\begin{equation}
  V_{\rm LoS}(\phi) = \sin i\,\cos\phi\,\sum_m V_m \cos[m(\phi-\phi_m)] \label{EqRingProj}
\end{equation}
where, as usual, $i$ is the inclination angle and we have assumed that the position angle of the major axis of the projected circle (ellipse) is at $\phi=0$. This may also\footnote{$V_{\rm LoS}(\phi)=\sin i \cos \phi \sum_m a_m \sin(m\phi) + b_m \cos(m\phi)$, as used by, e.g., \citet{1997MNRAS.292..349S}, is also equivalent.} be written as:
\begin{multline}
  V_{\rm LoS}(\phi) = \sin i\,\sum_m \frac{V_m}{2}\left[\cos((m-1)\phi - m\phi_m)\right. \\
\left.+ \cos((m+1)\phi - m\phi_m)\right]
\end{multline}

For example, when projecting an inclined circle of radius $R$ with average azimuthal velocity $V_0$, perturbed by an $m=2$ pattern with amplitude $V_2$ and phase $\phi_2$, the line-of-sight velocity along the resulting ellipse will be 
\begin{equation}
  V(\phi) = V_0\sin(i)\cos(\phi)\left[1 + \frac{V_2}{V_0}\cos(2(\phi-\phi_2))\right] \label{EqRingProj2}
\end{equation}
where we have assumed that the position angle of the major axis of the ellipse is at $\phi=0$. 

When the maxima of the mode are aligned with the major axis of the projection (i.e., $\phi_2=0^{\circ}$ or $180^{\circ}$), then the modulation increases the inferred rotation velocity. When $\phi_2=90^{\circ}$ or $270^{\circ}$, on the other hand, the projected kinematic major axis lines up with the minima and the inferred rotation velocities decrease.  

\begin{figure*}
  {\leavevmode \includegraphics[width=2\columnwidth]{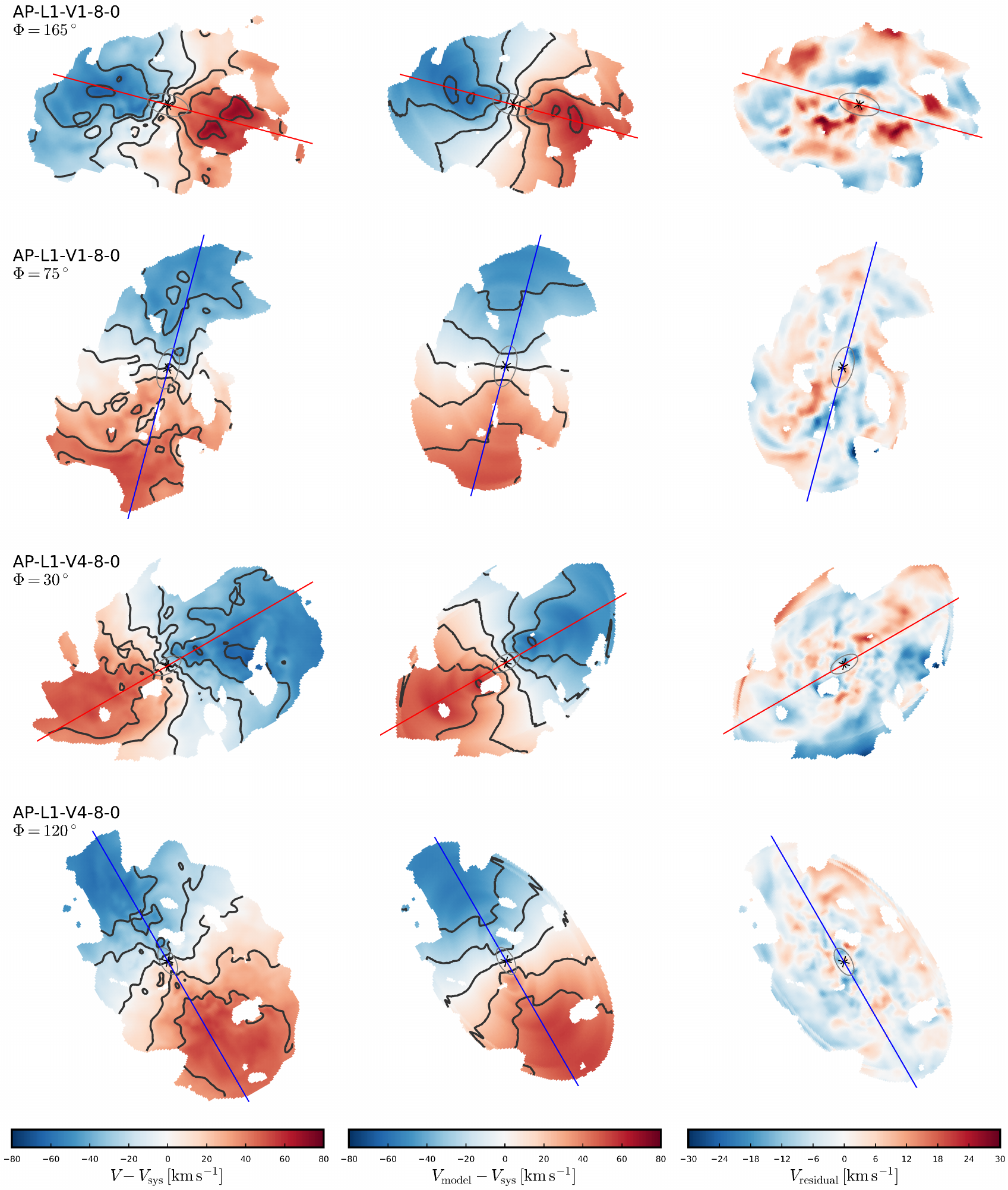}}
  \caption{\emph{Left column: }Velocity maps for the same two galaxies shown in Figs.~\ref{fig_barolo_fits}--\ref{fig_az_rot} along lines of sight which place the red (rows 1 \& 3, rotation curve systematically overestimated) or blue (rows 2 \& 4, rotation curve systematically underestimated) lines from the upper panels of Fig.~\ref{fig_az_dep} along the major axis. The grey ellipse marks $R=2\,{\rm kpc}$; the isovelocity contours are drawn at the same positions as the tick marks on the colour bars. \emph{Centre column: }Velocity maps extracted from the \barolo\ model data cubes for the same galaxies and orientations. \emph{Right column: }Difference of the left and centre columns (note that the colour scale is compressed).\label{fig_vfields}}
\end{figure*}

More generally, the velocity fluctuation about what is expected from uniform circular motion may be expressed by subtracting $V_0\sin(i)\cos(\phi)$  from Eq.~\ref{EqRingProj2},
\begin{equation}
 \Delta V_{\rm LoS}(\phi) = V_2\sin(i)\cos(\phi)\cos(2(\phi-\phi_2)),
\label{EqRingProjResid}
\end{equation}
which may also be written as
\begin{equation}
 \Delta V_{\rm LoS}(\phi) 
  = \frac{V_2}{2}\sin(i)\left[\cos(3\phi-2\phi_2) + \cos(\phi-2\phi_2)\right].
\label{EqRingProjResid2}
\end{equation}
In other words, in projection, a bisymmetric perturbation would be seen as simultaneous $m=1$ and $m=3$ perturbation to the line-of-sight velocities. The more visually obvious of the two is the three-peaked $m=3$ component. Such a three-peaked modulation may be quite difficult to detect in residual maps from tilted-ring models, as we show in Fig.~\ref{fig_vfields}. Here we show the same projections of the two galaxies from Fig.~\ref{fig_barolo_fits}. The left column shows the line-of-sight velocity map, the middle column the $1^{\rm st}$~moment of the \barolo\ map, and the residuals from the difference between these two are shown in the rightmost column.

The expected three-peaked pattern in the residuals is not readily apparent in any of the four cases illustrated. There are a number of reasons for this. First, the amplitude of the pattern is not very large ($\sim 15\,{\rm km}\,{\rm s}^{-1}$ on average according to Fig.~\ref{fig_az_dep}) and therefore only comparable to the rotation speed near the centre. Second, other residuals, not necessarily caused by the bisymmetric mode, dominate in the outer regions, obscuring the effect. Third, there are other harmonic modes in the velocity field which hinder a straightforward interpretation of the line-of-sight velocity field. For instance, there is a $m=2$ symmetric modulation of the radial velocities, as expected for gas orbits in a bar-like potential \citep[e.g.][]{2007ApJ...664..204S}, which tends to partially cancel the projected signature of the $m=2$ term in the azimuthal harmonic expansion. Finally, the tilted ring model attempts to provide a `best fit' by varying all available parameters so as to  minimize the residuals. Given the number of parameters available (each ring has, in principle, independent velocity, inclination, dispersion, and position angle), the resulting residuals are quite small, masking the expected three-peaked pattern, except perhaps in the most obvious cases.

We have attempted to measure the amplitudes and phases of the $m=2$ azimuthal harmonic perturbations seen in Fig.~\ref{fig_az_dep} from the line-of-sight velocity fields. However, all harmonic terms contribute to the line of sight velocities and must therefore in principle be modeled. Even a harmonic expansion fit up to only order $m=2$ is a nine-parameter problem prone to degeneracies: radial and azimuthal amplitudes and phases for $m=1$ and $m=2$, and the $m=0$ amplitude (circular velocity) must be determined, even when assuming (probably incorrectly) that vertical motions in the disc are negligible. In particular, the $m=1$ amplitude is degenerate with a combination of the systemic velocity and centroid; the radial and azimuthal terms of the same order are also degenerate given freedom in the phase; and the $m=2$ amplitude is partially degenerate with the inclination. 

Breaking these degeneracies requires strong assumptions, e.g. regarding the relative phases of the various terms in the expansion, or that the gas orbits form closed loops, which are difficult to justify based on 'observable' information. Ultimately, we find that we are unable to accurately recover the harmonic modes present in the discs of \apostle\ galaxies from their line of sight velocity maps without recourse to information which would be unavailable for real galaxies.

In spite of this complication, the azimuthal $m=2$ term clearly dominates in the case of \apostle\ galaxies, and is largely responsible for setting the recovered rotation velocities. Although the $m=1$ term in the azimuthal expansion sometimes has an amplitude comparable to the $m=2$ term, its degeneracy with the systemic velocity implies that in practice the effect on the recovered rotation curve is dominated by the $m=2$ harmonic. This is seen in Fig.~\ref{fig_az_rot}, where the inferred rotation velocity fluctuates with orientation angle as expected from an $m=2$ modulation, with the same phase as that shown in Fig.~\ref{fig_az_dep}.  We have verified this empirically by using \barolo\ to fit simple analytic models of differentially rotating discs with harmonic perturbations to their velocity fields. For independent $m=1$ and $m=2$ perturbations of the same amplitude, the $m=2$ symmetric perturbation always has a much stronger effect on the recovered rotation velocities, as measured by the average amplitude of their variation with phase angle.

\begin{figure*}
  {\leavevmode \includegraphics[width=1.5\columnwidth]{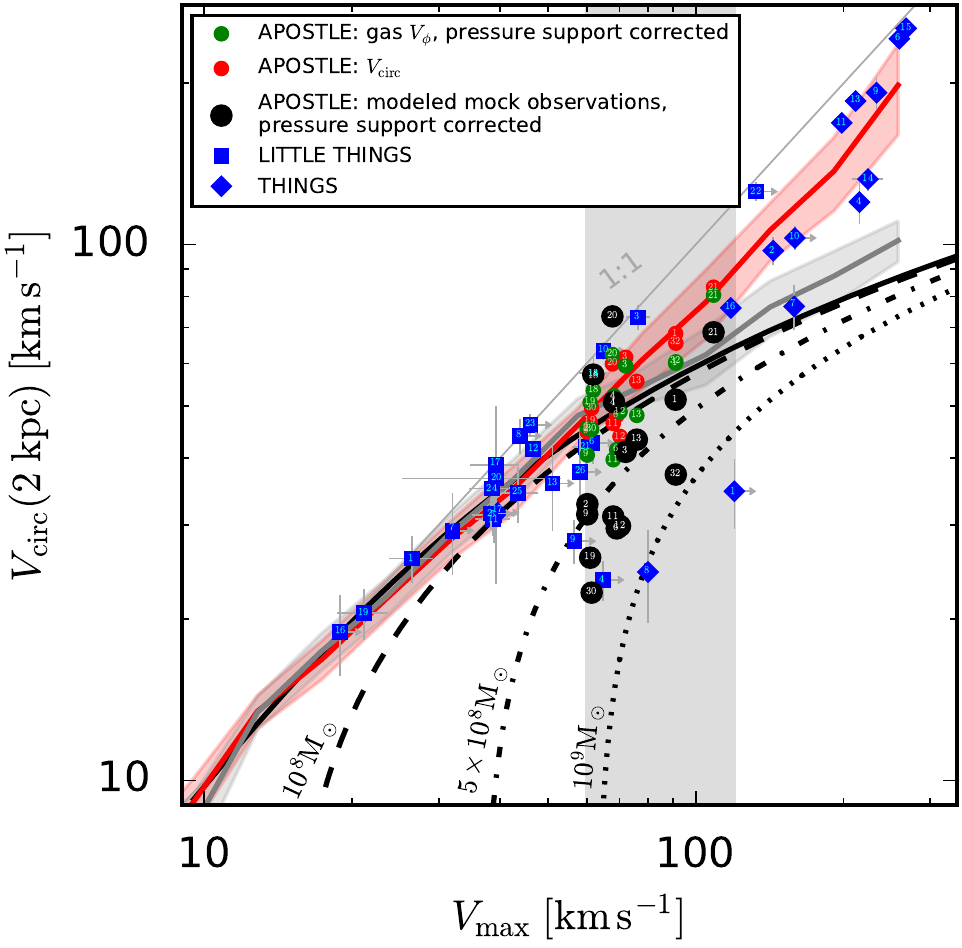}}
  \caption{Circular velocity at $2\,{\rm kpc}$ plotted against maximum circular velocity, a measure of the `central mass deficit'. The lines are reproduced from fig.~6 of \citet[][see their sec.~4.6 for additional details]{2015MNRAS.452.3650O}. The solid black line indicates the expected correlation for an NFW \citep{1996ApJ...462..563N,1997ApJ...490..493N} mass profile and the mass concentration relation of \citet{2014MNRAS.441..378L}; this is well traced by haloes in dark-matter-only versions of the \apostle\ and \eagle\ simulations. Values measured from the circular velocity profiles of galaxies from the (hydrodynamical) \apostle\ and \eagle\ simulations lie along the red line. The broken lines indicate the correlation for the same (NFW) profile but removing a fixed amount of mass from the central $2\,{\rm kpc}$, as labelled. Three points are shown for each of the \apostle\ galaxies in our sample: one for the circular velocity at $2\,{\rm kpc}$ (red), another for the gas azimuthal speed corrected for pressure support (green), and a third for the \barolo\ estimates of the rotation speed corrected for pressure support (black). Each quantity is shown as a function of the maximum of the circular velocity of each system, $V_{\rm max}$. Galaxies that we have flagged as kinematically disturbed (see Fig.~\ref{fig_profiles} and Sec.~\ref{SubSecGasRotVel}) are omitted. The grey shaded area marks our selection in $V_{\rm max}$, as in Fig.~\ref{fig_size_mass}. We show measurements from the \things\ (blue diamonds) and \littlethings\ (blue squares) surveys for comparison.\label{fig_v2_vmax}}
\end{figure*}

\begin{figure*}
  {\leavevmode \includegraphics[width=2\columnwidth]{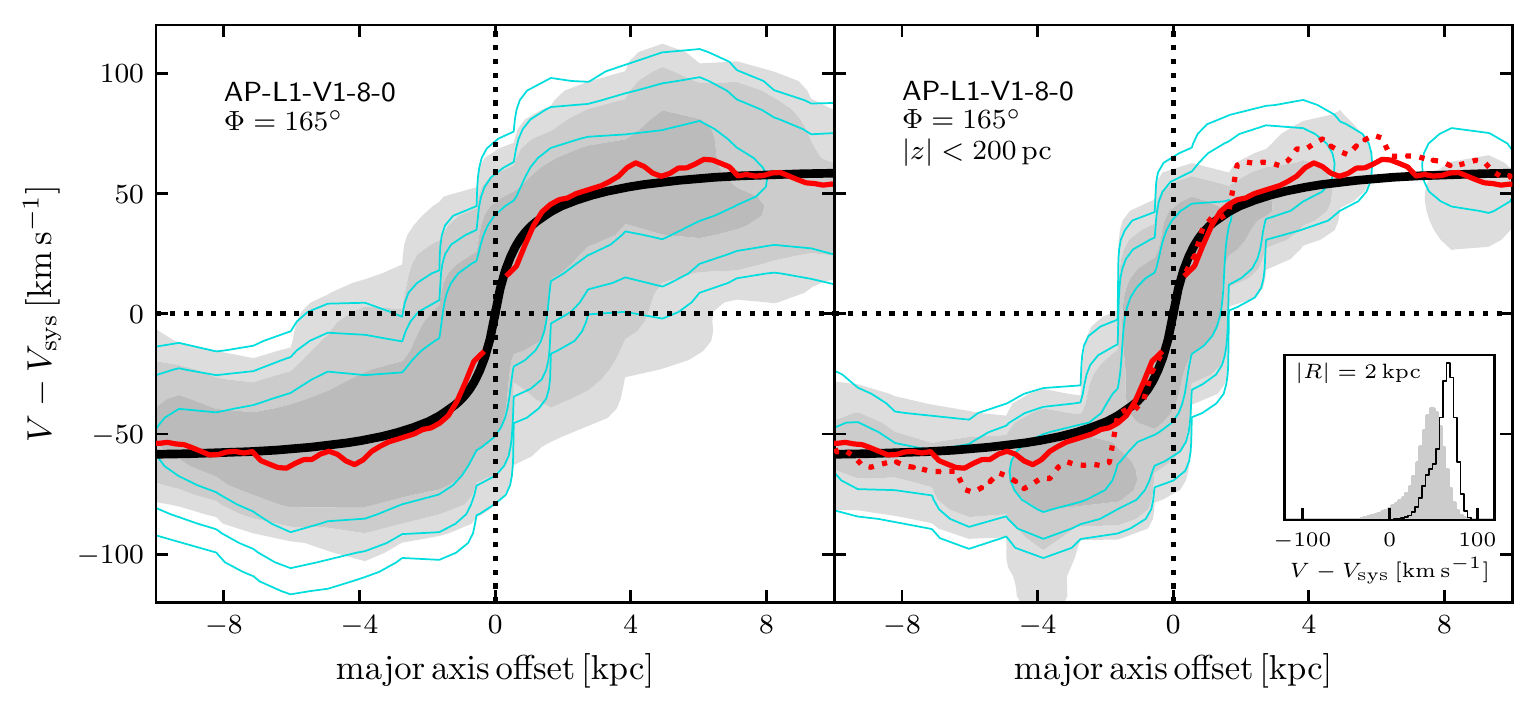}}\\
  {\leavevmode \includegraphics[width=2\columnwidth]{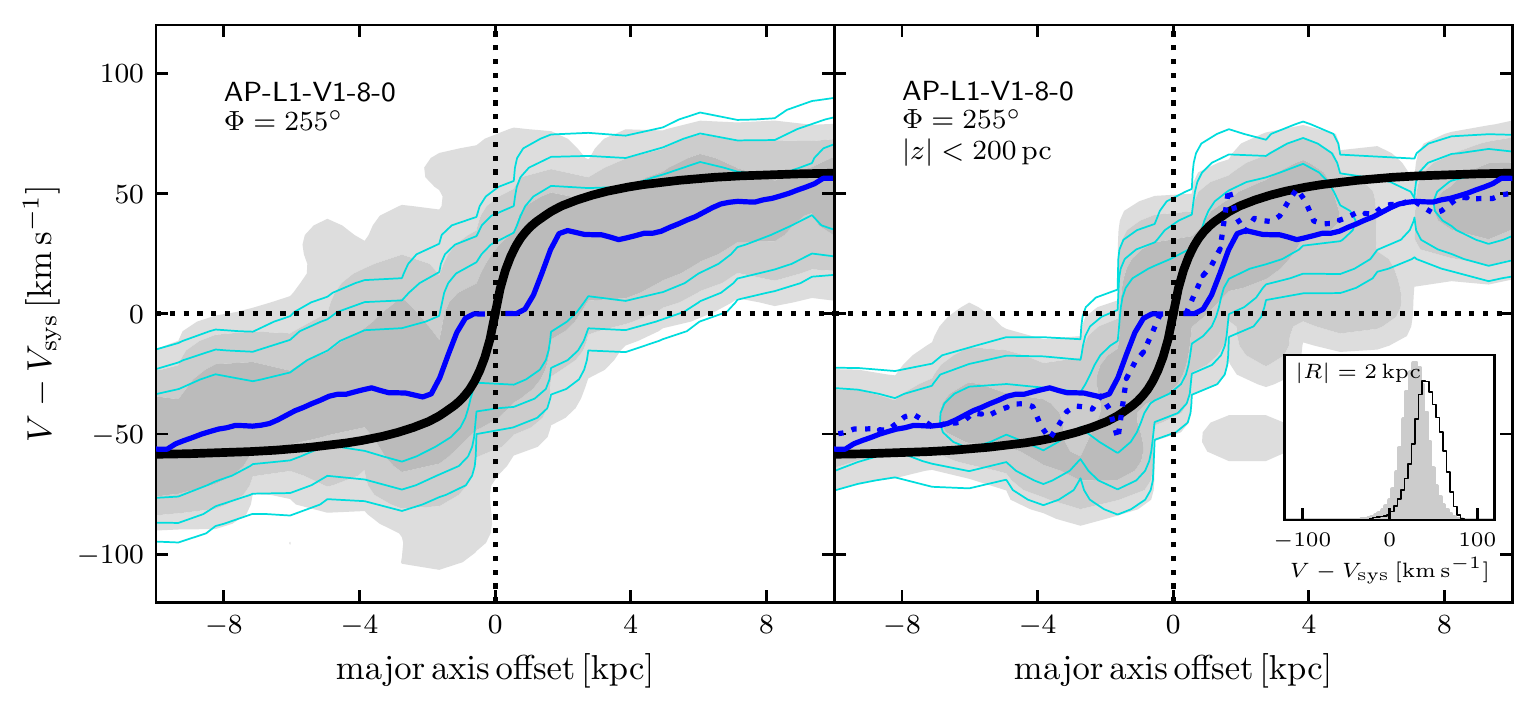}}
  \caption{\emph{Left panels: }Position-velocity (PV) diagrams along the major axis for two projections of \mbox{AP-L1-V1-8-0} offset by $90^\circ$: $\Phi=165^\circ$ (top row) and $\Phi=255^\circ$ (bottom row). The synthetically observed data for a $\sim 500\,{\rm pc}$ ($30\,{\rm arcsec}$) wide slice along the major axis is shown with filled contours; the levels are drawn at signal-to-noise levels of 0.3, 3 and 30, assuming a fiducial noise level of $0.4\,{\rm mJy}\,{\rm beam}^{-1}$, similar to that of the \things\ survey \citep{2008AJ....136.2563W}. The open contours correspond to the same slice through the model data cube produced by \barolo, with the contours drawn at the same levels. The coloured curve shows the \barolo-derived rotation curve, and the black curve shows the spherically-averaged circular velocity profile. \emph{Right panels: }Similar to the left panels, except that the synthetic observations have been made considering only those gas particles within $200\,{\rm pc}$ of the galactic mid-plane. The contour levels have been shifted 1-dex fainter to make the distribution visible. The coloured solid lines reproduce the rotation curves from the left panels, while the dotted lines show the rotation curve fit for the mid-plane gas. \emph{Inset panels: }Normalized histograms showing the velocity distributions along the major axis at $R=2\,{\rm kpc}$ in the synthetic data cubes, i.e. a `vertical' slice through the distributions shown with filled contours in the PV diagrams. The filled histogram corresponds to the entire gas distribution, the open histogram to only the gas within $200\,{\rm pc}$ of the midplane.\label{fig_pv_thick_thin}}
\end{figure*}

\subsection{Non-circular motions and the inner mass deficit problem}
\label{SecIMD}

The discussion above shows that non-circular motions can substantially affect the rotation curves of \apostle\ galaxies inferred from tilted-ring models. Because of their ubiquity, these motions can affect the inferred inner matter content of \apostle\ galaxies in a manner relevant to the `inner mass deficits' discussed in Sec.~\ref{SecIntro}.

Following \citet{2015MNRAS.452.3650O}, we estimate the inner mass deficit (a proxy for the importance of a putative `core') of a galaxy by comparing the observed and simulated relations between $V_{\rm circ}(2\, {\rm kpc})$ and the maximum circular velocity of the system, $V_{\rm max}$. The latter can usually be accurately determined because it is reached in the outer regions, which are less affected\footnote{The overall inclination of the system may in some cases be the ultimate impediment for accurate estimates of the rotation speed, especially when derived from kinematics alone \citep[see, e.g.,][for a discussion of a few examples]{2016MNRAS.460.3610O}.} by observational uncertainties. 

A lower limit for $V_{\rm circ}(2\, {\rm kpc})$ can be derived solely from the dark matter enclosed within $2\,{\rm kpc}$ for a $\Lambda$CDM halo of given $V_{\rm max}$, and is indicated by the grey band in Fig.~\ref{fig_v2_vmax}.  In the same figure, the red band indicates the $V_{\rm circ}(2\,{\rm kpc})$--$V_{\rm max}$ relation for all galaxies in the \apostle\ and \eagle\ suites of cosmological hydrodynamical simulations, including scatter. At low circular velocities the red and grey bands overlap, indicating that most low-mass \eagle\ and \apostle\ galaxies are dark matter dominated\footnote{Note that at the very low velocity end, $V_{\rm max}\lesssim 30\,{\rm km}\,{\rm s}^{-1}$, the maximum circular velocity is reached at radii close to $2$ kpc, so that $V_{\rm max} \approx V_{\rm circ}$(2 kpc).}. Note the small scatter in the $V_{\rm max}$--$V_{\rm circ}(2\,{\rm kpc})$ relation expected from these simulations. 

The red circles indicate the `equilibrium' \apostle\ galaxies selected for the present study, where $V_{\rm circ}(2\, {\rm kpc})$ and $V_{\rm max}$ are estimated directly from the mass profile as derived from the particle data. Green circles in Fig.~\ref{fig_v2_vmax} indicate the {\it average azimuthal velocities} at $2\,{\rm kpc}$ of the gas for the same \apostle\ sample. The good agreement between red and green circles is not surprising and simply reflects our definition of `equilibrium' as cases where the difference between average rotation speed and circular velocity at $2\,{\rm kpc}$ is smaller than $15$~per~cent. 

Blue symbols indicate the results for \things\ and \littlethings\ galaxies collated directly from the literature \citep{2008AJ....136.2648D,2011AJ....141..193O,2015AJ....149..180O}. Note that a number of galaxies lie below the red band; these are galaxies with an apparent `inner deficit' of mass at that radius (i.e., `cores') compared with the cold dark matter prediction. The broken lines are labelled by an estimate of this mass deficit, expressed in solar masses. Galaxies like DDO~47 (blue square `4') and DDO~87 (blue square `9') lie well below the expected relation; they are clear examples of galaxies with the slowly rising rotation curves traditionally associated with `cores' in the dark matter (Fig.~\ref{fig_rotcurves}).

The black circles in Fig.~\ref{fig_v2_vmax}, finally, indicate rotation speeds at $2\,{\rm kpc}$ for all `equilibrium' \apostle\ galaxies, derived using \barolo\ for a fixed inclination ($i=60^\circ$) and a single random orientation per galaxy. The obvious difference between black and green circles highlights two important conclusions. One is that non-circular motions in the inner regions are important enough to produce, at times, deviations from the expected relation as large as measured in observed galaxies. The second is that $V_{\rm rot}$ generally underestimates $V_{\rm circ}$ at $2\,{\rm kpc}$. Overestimates also occur in some cases, but these are rare and usually milder. 

This indicates that the discrepancy between inner rotation and circular speeds is not solely a result of a bisymmetric modulation of the velocity field, where overestimates should occur as frequently as underestimates. The systematic underestimate of the circular velocity must arise from other effects, such as (i) the non-negligible thickness of the gas disc (which causes gas at different radii and heights to fall along the line of sight; see below); (ii) morphological irregularities that may push the gas temporarily out of equilibrium \citep[e.g. \HI\ bubbles, see also][]{2016MNRAS.462.3628R,2017A&A...607A..13V}; and (iii) underestimated `pressure' support from random motions in the gas \citep{2017MNRAS.466...63P}. Our main conclusion is that tilted-ring modelling of \apostle\ galaxies results in a diversity of inner rotation curve shapes and {\it apparent} inner mass deficits that are comparable to those of observed galaxies, mainly due to non-circular motions in the gas.

\subsection{Disc thickness and projection effects}
\label{SecDiscThickProj}

Of the three possible explanations for the systematic underestimate of the circular velocity at $2\,{\rm kpc}$, the thickness of the gaseous discs seems to be the leading factor in \apostle\ galaxies. We show this in Fig.~\ref{fig_pv_thick_thin}, using as an example \mbox{AP-L1-V1-8-0}, whose \barolo\ rotation curves for two orthogonal orientations are shown in the top-left panel of Fig.~\ref{fig_barolo_fits}. As discussed above (Sec.~\ref{subsec_ncm_orientation}), the blue curve ($\Phi=255^\circ$)\footnote{We show here the orientation offset $180^\circ$ from that in Fig.~\ref{fig_barolo_fits} -- the effect of the bisymmetric perturbation is much the same, but by chance this orientation illustrates the effect of the thick disc more clearly than the $\Phi=75^\circ$ orientation.} subtantially {\it underestimates} $V_{\rm circ}(r)$ because the kinematic major axis of the projection coincides with the minima of the $m=2$ perturbation pattern shown in Fig.~\ref{fig_az_dep}. Why doesn't then the red curve ($\Phi=165^\circ$), where the kinematic major axis traces the maxima of the $m=2$ pattern, {\it overestimate} $V_{circ}$ by a similar amount?

The answer may be gleaned from Fig.~\ref{fig_pv_thick_thin}, where we show `position-velocity' (PV) diagrams (i.e., the distribution of line-of-sight velocities along the kinematic major axis) for this galaxy. Red and blue curves are the rotation velocites estimated by \barolo; black curves indicate the true circular velocity on the plane of the disc. The left panels show the PV diagrams including all the \HI\ gas in the galaxy; the panels on the right, on the other hand, only include gas very close  to the disc plane  (i.e., $|z|<200\,{\rm pc}$)\footnote{This height was chosen as the minimum height for which a synthetic observation could be constructed without becoming unduly dominated by shot noise in the simulation; the precise value is not otherwise significant.}. The right-hand panels exclude extra-planar gas, which tends to rotate more slowly and to lower the average speed at given $R$. When only gas near the plane is considered, the inferred rotation velocities do indeed under and overestimate the true circular velocity by similar amounts, as expected for a bisymmetric perturbation. The effect of the extraplanar gas is to bring down average speeds in both cases, effectively cancelling the overestimate in the case of $\Phi=165^\circ$, and leading to systematic underestimation of the inner circular velocities when averaged over all orientations. 

\begin{figure}
  {\leavevmode \includegraphics[width=\columnwidth]{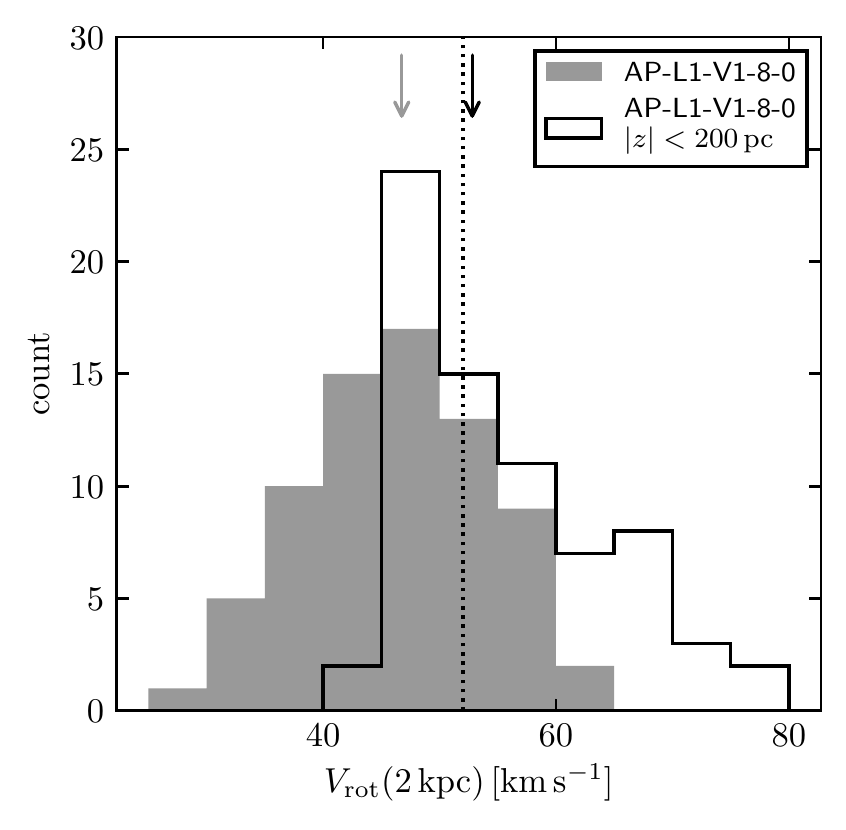}}
  \caption{Distribution of \HI\ rotation velocity at $R=2\,{\rm kpc}$ as recovered using \barolo\ for \mbox{AP-L1-V1-8-0} for 72 orientations, all at $i=60^\circ$, each separated in viewing angle by $5^\circ$ in azimuth. The filled histogram corresponds to synthetic data cubes constructed using the entire \HI\ gas distribution, whereas for the open histogram, data cubes where any gas more than $200\,{\rm pc}$ from the galactic mid-plane has been discarded were used instead. The medians of the two distributions are marked with the arrows of corresponding colour, and the circular velocity at $2\,{\rm kpc}$ is marked by the vertical dotted line.\label{fig_hist_thick_thin}}
\end{figure}

This is shown explicitly in Fig.~\ref{fig_hist_thick_thin} (see also inset panels in Fig.~\ref{fig_pv_thick_thin}), where we plot the distribution of \barolo\ inferred rotation velocities at $2\,{\rm kpc}$ for $72$ different orientations of the galaxy, each separated by $5^\circ$ in azimuth. The filled histogram correspond to all \HI\ gas, the open histogram only to gas with $|z|<200\,{\rm pc}$. The combined effects of non-circular motions and extraplanar gas lead to a relatively large dispersion in the estimated values (of order $8.3\,{\rm km}\,{\rm s}^{-1}$ and $7.7\,{\rm km}\,{\rm s}^{-1}$ for the open and filled histograms, respectively), as well as a substantial shift in the median value; from  $52.8\,{\rm km}\,{\rm s}^{-1}$ to  $46.7\,{\rm km}\,{\rm s}^{-1}$ (for reference, the circular velocity at this radius is $52.0\,{\rm km}\,{\rm s}^{-1}$). 

\begin{figure*}
  {\leavevmode \includegraphics[width=2\columnwidth]{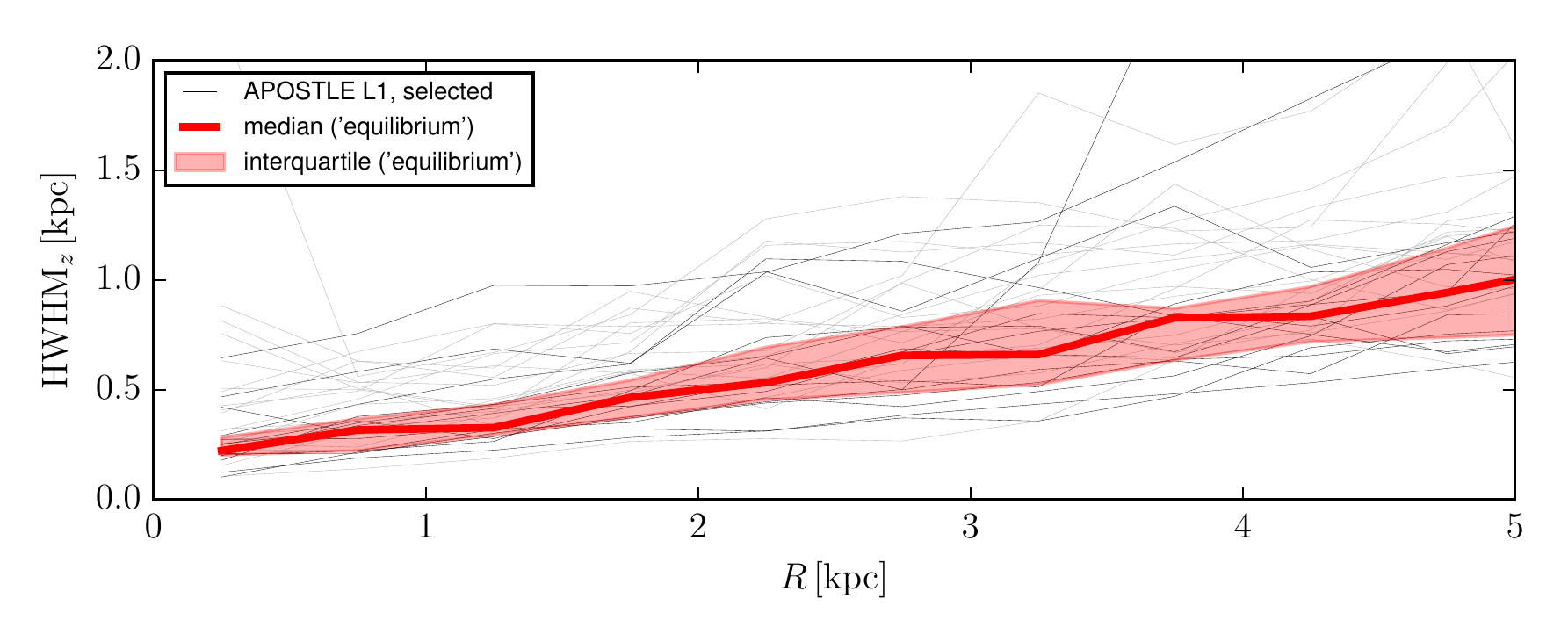}}
  \caption{Height enclosing half the mass in \HI\ as a function of radius for \apostle\ galaxies in our sample (thin lines). Our `equilibrium' (`non-equilibrium') subsample is shown with black (grey) lines. `Equilibrium' galaxies tend to be slightly thinner, on average. The median and interquartile interval of the distribution of `equilibrium' galaxies are shown by the heavy red line and shaded area, respectively.\label{fig_heights}}
\end{figure*}

We end this discussion by presenting, in Fig.~\ref{fig_heights}, the thickness of \apostle\ \HI\ discs. Here we show, as a function of cylindrical radius, the height ${\rm HWHM}_z$ containing half the \HI\ mass. Black and grey thin lines indicate our sample of `equilibrium' and `non-equilibrium' galaxies, respectively. The thick red line and shaded areas highlight the median and interquartile half-mass heights for the `equilibrium' sample. Note that `equilibrium' galaxies are slightly thinner, on average; their typical half-HI height is $\sim 500$ pc at $R=2\,{\rm kpc}$. \apostle\ discs also flare outwards \citep{2018MNRAS.473.1019B}; the typical \HI\ half-height climbs to $\sim 1\,{\rm kpc}$ at $R=5\,{\rm kpc}$.

\section{Inner mass deficits of observed galaxies}
\label{SecIMDObs}

We explore next whether non-circular motions and thick, differentially rotating discs seen in projection might be at least partially responsible for the observed `inner mass deficits' seen in Fig.~\ref{fig_v2_vmax}. This is a lengthy topic that requires a detailed individual study of each galaxy, which we plan to pursue in future work. Our main purpose here is to provide a `proof of concept' that at least some of the galaxies with large inferred deficits/cores might indeed have rotation curves that are unduly affected by model shortcomings similar to those seen in the analysis of \apostle\ galaxies.

\subsection{Non-circular motions}

We motivate the potential applicability of our findings from Sec.~\ref{subsec_ncm_orientation} to real galaxies by considering two example systems. DDO~47 and DDO~87 have reported inner rotation speeds which are well below those expected for CDM haloes (see blue squares labelled `4' and `9' in Fig.~\ref{fig_v2_vmax}). These galaxies have been modelled with \barolo\ by \citet{2017MNRAS.466.4159I}, and we adopt here the same parameters as in that study. The \barolo\ rotation curves are consistent (within the reported errors) with those of \citet{2015AJ....149..180O}, despite significant differences in the algorithms implemented in the codes used in each case.

\begin{figure*}
  {\leavevmode \includegraphics[width=2\columnwidth]{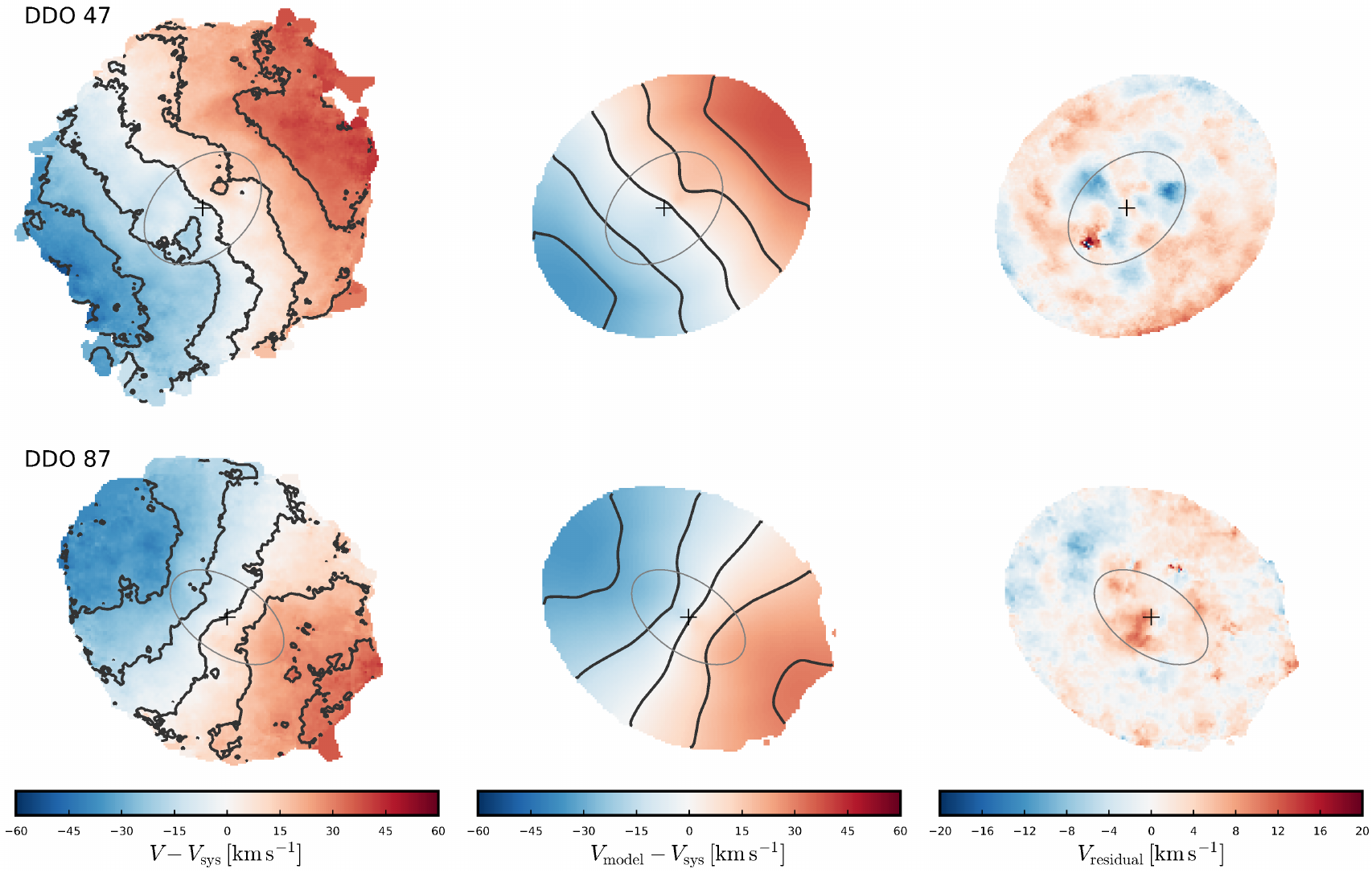}}
  \caption{\emph{Left column: }Velocity maps for the \littlethings\ galaxies DDO~47 and DDO~87, which have slowly rising rotation curves (Fig.~\ref{fig_rotcurves}). The grey ellipse marks $2.5\,{\rm kpc}$; the isovelocity contours are drawn at the same positions as the tick marks on the colour bars. \emph{Centre column: }Velocity maps extracted from \barolo\ model data cubes for the same galaxies. We use the same configuration for \barolo\ as in \citet{2017MNRAS.466.4159I}, and confirm that we recover the same rotation curves. \emph{Right column: }Difference of the left and centre columns, notice the `3 petal' pattern near the centre.\label{fig_vfields_obs}}
\end{figure*}

The $1^{\rm st}$~moment maps of DDO~47 and DDO~87 are shown in the left column of Fig.~\ref{fig_vfields_obs}. The \barolo\ velocity fields are shown in the middle, and the residuals in the panels on the right. Although, as discussed in Sec.~\ref{SubsecObsApplicability}, we are unable to accurately constrain the parameters of a harmonic expansion from the line of sight velocity maps, the clear three-peaked (`clover leaf') pattern in the residuals (with amplitudes of order $\sim 10\,{\rm km}\,{\rm s}^{-1}$; about half the inferred rotation velocity at $2\,{\rm kpc}$) strongly suggests the presence of bisymmetric kinematic modulations near the centres of these galaxies. The phase of the residual patterns -- with maxima (red) lying approximately along the major axis on the approaching side of the disc (blue in the left column; see Fig.~\ref{fig_schematic} for a schematic) -- further suggests that the recovered rotation velocities might very well underestimate the true inner circular velocity. Although a definitive verdict on whether these galaxies are actually consistent with cuspy CDM haloes must await a more detailed analysis, we regard the evidence for non-circular motions shown in the right-hand panels of Fig.~\ref{fig_vfields_obs} as strong enough to call into question the conclusion that large `cores' are present in these galaxies.

\subsection{Disc thickness and projection effects}

We argued in Sec.~\ref{SecDiscThickProj}  that the finite thickness of \apostle\ discs is mainly responsible for the systematic underestimation of inner circular velocities shown in Fig.~\ref{fig_v2_vmax}. Could this explanation also hold for observed galaxies? 

We begin by noting that assessing the impact of disc thickness on the inferred rotation curve is challenging, since it depends not only on the gas scale height and its radial dependence, but also on the gas velocity gradient in both the radial and vertical direction, none of which are entirely straightforward to constrain in real galaxies. We defer a more detailed analysis to a future study, and will only attempt here a preliminary comparison between the thickness of observed and \apostle\ discs to motivate the fact that our conclusion in Sec.~\ref{SecDiscThickProj} may indeed be applicable to real galaxies.

\begin{figure}
  {\leavevmode \includegraphics[width=\columnwidth]{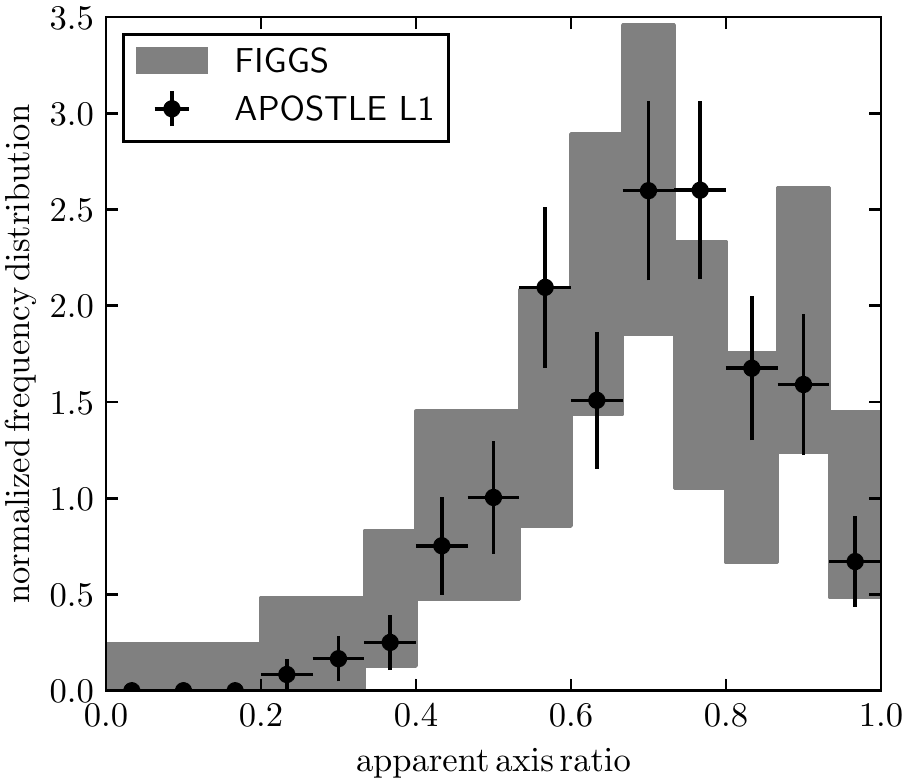}}
  \caption{Distribution of apparent beam-corrected axis ratios of the \HI\ distribution in \apostle\ galaxies (points with error bars) at a column density of $10^{19}\,{\rm cm}^{-2}$. The sample is drawn from $\sim 200$ \apostle\ dwarfs with $20 < V_{\rm max}/{\rm km}\,{\rm s}^{-1} < 120$ and is constructed to approximately match the median and width of the \HI\ mass distribution of the FIGGS sample of \citet[][see also \citealp{2008MNRAS.386.1667B}]{2010MNRAS.404L..60R}, whose measurement is reproduced here (grey band). Each \apostle\ galaxy is viewed from a random angle and convolved with a $1\,{\rm kpc}$ (FWHM) beam, similar to the $44\,{\rm arcsec}$ beam used to observe the FIGGS sample, at the median distance of $5\,{\rm Mpc}$ of those galaxies. The vertical error bars/shaded areas represent the Poisson errors on the measurements; the horizontal error bars/shaded areas are the bin widths.\label{fig_axisratio}}
\end{figure}

A simple (but statistical) measure of disc thickness is provided by the distribution of apparent axis ratios (i.e., projected shapes) of \HI\ discs. This has been reported for FIGGS\footnote{Faint Irregulars Galaxies GMRT Survey.} \citep{2008MNRAS.386.1667B} galaxies by \citet{2010MNRAS.404L..60R}, whose results we reproduce in Fig.~\ref{fig_axisratio} (shaded histogram). The symbols with error bars in the same figure correspond to a sample of \apostle\ discs chosen to match the median and width of the \HI\ mass distribution of FIGGS galaxies. (Note that this is not the same sample we have used for the kinematic analysis in previous sections.) The two distributions are in very good agreement, suggesting that FIGGS and \apostle\ discs have similar thicknesses, at least according to this statistic.

Direct estimates of the half-mass thickness of  \HI\ discs are available only for small samples of galaxies,  and are subject to selection biases and observational uncertainties that hinder a proper comparison with the \apostle\ results presented in Fig.~\ref{fig_heights}. For example, edge-on galaxies are good targets in principle, but they are typically identified from samples of thin galaxies, thereby possibly incurring selection  biases difficult to quantify and correct for. 

\citet{2010A&A...515A..62O} analyze a sample of 8~\HI-rich, approximately edge-on late-type galaxies and report  ${\rm HWHM}_z$ in the range $0.3$--$1.0\,{\rm kpc}$ at a radius of $5\,{\rm kpc}$ (see their fig.~24; note that they show ${\rm FWHM}_z=2\times {\rm HWHM}_z$). \cite{2017MNRAS.464...32P} re-analyzed a subset of the same galaxies and also report estimates of disc thickness in the central regions that are consistent with those from \citet{2010A&A...515A..62O}. These scale heights are on average slightly thinner than the \apostle\ equilibrium sample, where we find a half-mass \HI\ height of $1.0\pm 0.2$ at $5\,{\rm kpc}$, but this is perhaps not unexpected given the caveat expressed in the preceding paragraph.

Another set of estimates of disc thickness for dwarf galaxies comparable to our \apostle\ sample (DDO~154, Ho~II, IC~2574 and NGC~2366), is provided by \citet{2011MNRAS.415..687B}. These authors use a dynamical model that requires assumptions about geometry and about the contribution of stars, gas, and dark matter to the gravitational potential. They report ${\rm HWHM}_z(5\,{\rm kpc})$ in the range $0.2$--$0.8\,{\rm kpc}$, or a factor of $\sim 2$ thinner than the `equilibrium' \apostle\ galaxies at the same radius.

Finally, an estimate of the importance of extra-planar gas can also be made by integrating the amount of `kinematically anomalous' gas in a galaxy, which refers to the `non-Gaussian' tail of the \HI\ velocity distribution at each pixel \citep{2002AJ....123.3124F,2008A&ARv..15..189S}. When applied to our projected \apostle\ velocity fields, we find that between $10$ and $20$~per~cent of the \HI\ flux of a galaxy may be classified as `kinemaically anomalous'. This is actually in good agreement with the results of the same procedure applied to \littlethings\ galaxies, although a quantitative comparison is not straightforward because it depends on observational issues such as the treatment of noise, smoothing, masking, etc., which are difficult to replicate in simulations. Interestingly, we find that, in APOSTLE, a fair fraction of the `kinematically anomalous' gas is actually not extra-planar, but, rather, gas close to the mid-plane that has been disturbed by recent episodes of star formation, or other kinematic perturbations. Again, a detailed comparison between observation and simulation requires a more meticulous study than the preliminary exploration we attempt here, but the overall agreement between observation and simulation for this measure is reassuring.

In conclusion, and taking these various measurements at face value, it appears that our sample of \apostle\ \HI\ discs are typically somewhat thicker at $R\lesssim 5\,{\rm kpc}$ than their observed counterparts. Whether this slight offset is enough to invalidate our conclusion that the finite thickness of gas may be responsible for the systematic `inner mass deficits' shown in Fig.~\ref{fig_v2_vmax} is unclear. Reaching a more definitive conclusion demands a more detailed analysis than warranted by the scope of the present paper. 

\subsection{Non-circular motions, surface brightness, and rotation curve slopes}

Finally, we consider whether non-circular motions are consistent with the existence of relatively tight correlations betwen the shape of the rotation cuves and structural properties of the galaxy, such as the inner surface density of gas and stars. Indeed, it has long been appreciated that high-surface brightness galaxies typically have faster rising inner rotation curves than lower surface brightness systems \citep{1996MNRAS.283...18D}. Could these correlations exist if the inner parts of rotation curves are substantially affected by non-circular motions?

We address this question by discussing the correlation between the central rotation curve slope, which we parametrize using the ratio between the radius at which the circular velocity curve reaches $90$~per~cent of its maximum value and the velocity at that radius\footnote{This measure has the advantage of being independent of galaxy size; \apostle\ dwarfs have systematically larger \HI\ (Fig.~\ref{fig_size_mass}) and stellar \citep{2017MNRAS.469.2335C} sizes than observed.}, $V(R_{90})/R_{90}$, and the central surface density of the stellar component, $\Sigma_\star(0)$.

\begin{figure}
  {\leavevmode \includegraphics[width=\columnwidth]{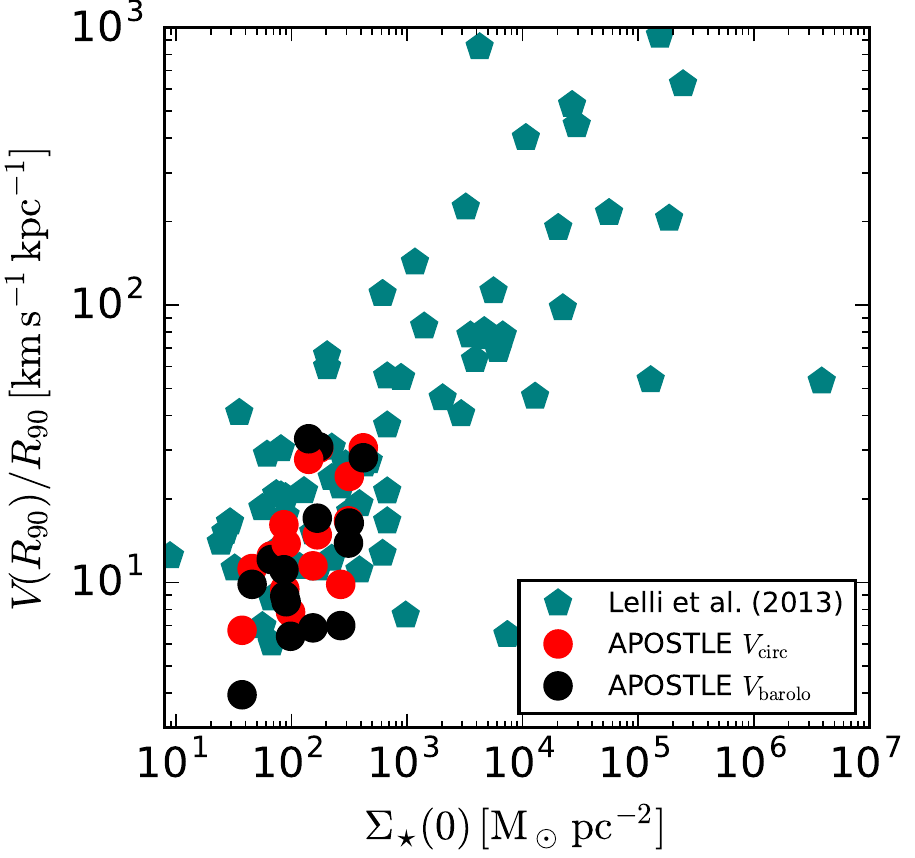}}
  \caption{Central rotation curve slope, which we parametrize as $V(R_{90})/R_{90}$, as a function of central stellar surface density, $\Sigma_\star(0)$; $R_{90}$ is the radius where the circular velocity curve reaches $90$~per~cent of its maximum amplitude. For our sample of `equilibrium' \apostle\ galaxies, we show both the central rotation curve slope as measured from the circular velocity curve (red points), and from the rotation curve derived using \barolo\ (black points). We show for comparison measurements from \citet[][pentagons]{2013MNRAS.433L..30L}. We have converted their central surface brightnesses to central surface densities assuming a mass-to-light ratio in the $R$-band of $\Upsilon_R^\star=3.0\,{\rm M}_\odot\,{\rm L}_\odot^{-1}$, and a Solar magnitude of $4.5$ in the $R$-band.\label{fig_innerslope}}
\end{figure}

We show the observed correlation based on measurements reported in the recent compilation of late type galaxies by \citet[][note that these authors use a different definition of central rotation curve slope]{2013MNRAS.433L..30L} and compare it with \apostle\ results in Fig.~\ref{fig_innerslope}. We make two measurements of $V(R_{90})/R_{90}$ for \apostle\ galaxies; one using the true circular velocities of the galaxies (red points), and a second one using the rotation curves recovered using \barolo\ (black points), for the same random orientations as used in Fig.~\ref{fig_v2_vmax}. $\Sigma_\star(0)$ is measured, as in the observations, by extrapolating a fit to the exponential portion of the radial stellar mass profile.

Fig.~\ref{fig_innerslope} shows good agreement between observations and \apostle. For the surface brightness range sampled by our sample of \apostle\ galaxies rotation curves rise in general fairly slowly, with typical slopes of order $\sim 10\,{\rm km}\,{\rm s}^{-1}\,{\rm kpc}^{-1}$. This is quite well reproduced in the \apostle\ sample, regardless of whether the true circular velocity curves or the  \barolo\ rotation curves are used. Fig.~\ref{fig_innerslope} suggests that there is no conflict between the slope-surface brightness correlations and our results, at least in the range explored by the \apostle\ sample.

\section{Summary and Conclusions}
\label{SecConc}

We analyse synthetic \HI\ observations of $33$ galaxies from the \apostle\ suite of $\Lambda$CDM cosmological hydrodynamical simulations of galaxy formation. Our sample includes all systems with maximum circular velocities in the range $60$--$120\,{\rm km}\,{\rm s}^{-1}$, simulated at the highest (L1) resolution available in \apostle. The \HI\ properties of these galaxies follow closely the scaling laws relating the size, mass and kinematics of observed galaxy discs.

We derive \HI\ rotation curves using the same kind of tilted-ring modelling usually applied to interferometric \HI\ observations.  We find that in many cases the rotation curves inferred by this modelling depend sensitively on the orientation of the chosen projection. Particularly affected are the rotation speeds in the inner regions of individual galaxies, which fluctuate systematically as the orientation angle of the line of sight is varied, at fixed inclination. The fluctuations are due to non-circular motions of the gas velocity field in the disc plane. Bisymmetric ($m=2$) variations in the azimuthal velocity of the gas have the most obvious effect, but other asymmetric patterns are also present.

Maximal deviations from the actual circular velocity are obtained when the line of nodes of the projection coincides with the principal axes of the $m=2$ pattern. In particular, rotation speeds can severely underestimate the circular velocity when the line of nodes is aligned with the minima of the modulation pattern. Conversely, alignment of the line of nodes with the maxima can result in a substantial overestimate of the rotation speed. The case of an underestimate is further exacerbated by lower velocity extra-planar gas along the line of sight, whereas in the case of an overestimate the two effects partially compensate each other. Those cases which result in a slowly-rising rotation curves would be erroneously interpreted as signaling a severe `inner mass deficit' compared with the $\Lambda$CDM predictions.

Slowly-rising rotation curves are often interpreted as evidence for a `core' (rather than a cusp) in the dark matter central density profile. This raises the possibility that some galaxies with `cores', and especially those where the inner mass deficit is extremely large, are galaxies whose inner rotation curves have been unduly affected by modelling uncertainties. In fact, all \apostle\ galaxies have central dark matter cusps that have been minimally affected, if at all, by the baryonic assembly of the galaxy.

Bisymmetric patterns map, in projection, into one- and three-peaked patterns in the residual velocity fields, but these are often difficult to detect in tilted ring model residuals, as they may be effectively masked by the numerous free parameters of model fits and by degeneracies which arise in projected coordinates. They are easily appreciated, however, in some galaxies that show extreme deviations from the steeply-rising rotation curves expected from the cuspy CDM halo density profiles, such as  DDO~47 and DDO~87.  The residual pattern in these galaxies suggests the presence of bisymmetric gas flows strong enough to substantially affect the kinematic modelling. Not only is the true average azimuthal speed of the gas, $V_{\rm rot}(R)$, very difficult to extract in this case, but even if that were possible, it may still differ substantially from the true circular velocity of the system. Galaxies such as these may well have a cuspy dark matter halo despite the apparent slow rise of their inner rotation curves.

Our results thus suggest a possible resolution of the rotation curve diversity problem highlighted by \citet{2015MNRAS.452.3650O}. The problem could, at least in part, result from the failure of tilted-ring models applied to discs of finite thickness with non-negligible non-circular motions. In this context, the systematic underestimation of inner circular velocities may just reflect the finite thickness of gas discs. 

It is therefore important to design simple and model-independent diagnostics of non-circular motions to identify the regions where tilted-ring model results are suspect and where uncertainties from such techniques are almost certainly underestimated. Non-circular motions should not be `assumed' to be unimportant, as is often done, but demonstrated to be so through careful analysis, keeping in mind that they may appear weaker than they are. Tighter constraints on the scale heights of the \HI\ discs of dwarfs would also be particularly useful. 

Much remains to be done to explore fully these ideas. One issue is the origin of the bisymmetric perturbations, which seem ubiquitous in \apostle, as well as their amplitude and radial dependence: are they radially-coherent modes of constant phase angle, as in a bar or in a triaxial halo potential, or of varying pitch angle, as in two-armed spirals? \citep[See][for a first attempt.]{2018MNRAS.476.2168M} Are they associated with disc instabilities, halo triaxiality, accretion events, or perhaps even triggered and sustained by the stochastic nature of star formation and feedback? Are non-circular motions in observed dwarfs as strong as they seem to be in our simulated galaxies? What are the implications for $\Lambda$CDM models?

Another issue that remains pending is a case-by-case demonstration that non-circular motions provide a viable explanation for galaxies with alleged `cores'. Although difficult to detect, we are hopeful that our models will provide enough guidance to devise effective methods for finding and characterizing them. Finally, what are the statistics of these modulations? What fraction of galaxies do they affect, and by how much? Is this consistent with available data? For simplicity, we have fixed the inclination of all galaxies in our sample to $60^\circ$; what is the result of adopting a more realistic distribution of inclination angles?

These are all important considerations that we plan to address in future contributions, where we hope to clarify whether the inner mass deficit problem, and the related cusp-core problem, results in whole or in part from an underestimate of the impact of non-circular motions on the kinematic modelling of disc galaxies, or else is a true predicament for the otherwise extremely successful $\Lambda$CDM paradigm.

\section*{Acknowledgements}\label{sec-acknowledgements}

KO and AM thank E. di Teodoro for technical assistance with \barolo\ and useful discussions. We thank S.-H.~Oh and E.~de~Blok for data contributions, and the \things, \littlethings\ and SPARC survey teams for making their data publicly available. We thank the anonymous referee whose comments helped to improve substantially the manuscript. This work was supported by the Science and Technology Facilities Council (grant number ST/F001166/1). CSF and ABL acknowledge support by the Science and Technology Facilities Council (grant number ST/L00075X/1). JS acknowledges ERC grant agreement 278594-GasAroundGalaxies and the Netherlands Organisation for Scientific Research (NWO) through VICI grant 639.043.409. This work used the DiRAC Data Centric system at Durham University, operated by the Institute for Computational Cosmology on behalf of the STFC DiRAC HPC Facility (www.dirac.ac.uk). This equipment was funded by BIS National E-infrastructure capital grant ST/K00042X/1, STFC capital grant ST/H008519/1, and STFC DiRAC Operations grant ST/K003267/1 and Durham University. DiRAC is part of the National E-Infrastructure. This research has made use of NASA's Astrophysics Data System.

\bibliography{paper}

\appendix
\section{Properties of simulated and observed galaxies}
\label{SecApp1}

In Tables~\ref{tab_APprops} \& \ref{tab_thingsprops} we collect some key properties of our sample of simulated galaxies and the \things\ \citep{2008AJ....136.2563W} and \littlethings\ \citep{2012AJ....144..134H} galaxies, respectively, including the key to the numeric labels used in Figs.~\ref{fig_size_mass}, \ref{fig_symmetry} \& \ref{fig_v2_vmax}.

Not all galaxies in these surveys are amenable to kinematic analysis. We select the \things\ galaxies analysed in \citet{2008AJ....136.2648D} and \citet{2011AJ....141..193O}. In cases where the same galaxy is analysed in both studies, we use the more recent analysis. We exclude some of the galaxies:
\begin{itemize}
\item NGC~4826 has two counter-rotating discs, making the interpretation of the rotation curve as a circular velocity curve highly uncertain.
\item Ho~II is the same galaxy as DDO~50, which appears in the \littlethings\ sample.
\item We omit DDO~53, DDO~154 and NGC~2366, which are included as part of the \littlethings\ sample.
\item Ho~I has a particularly low inclination of $\lesssim 14^\circ$ \citep{2011AJ....141..193O}, which makes the necessary inclination correction unacceptably large and uncertain.
\end{itemize}
We select all of the \littlethings\ galaxies analysed in \citet{2015AJ....149..180O}.

\begin{table*}
  \caption{Selected properties of the synthetically observed \apostle\ galaxies in our sample. Columns: {\bf (1)}~Number corresponding to labels in Figs.~\ref{fig_size_mass}, \ref{fig_symmetry} \& \ref{fig_v2_vmax}, entries in bold face are those referred to as `equilibrium' systems in the text and shown in Fig.~\ref{fig_v2_vmax}; {\bf (2)}~Object label as described in Sec.~\ref{SubsecApostle}; {\bf (3)}~Neutral hydrogen mass (the gas mass used in the BTFR is $M_{\rm bar} = M_\star + 1.4M_{\rm HI}$); {\bf (4)}~Stellar mass; {\bf (5)}~\HI\ size defined as the radius where $\Sigma_{\rm HI}$ drops to $1\,{\rm M}_\odot\,{\rm pc}^{-2}$, measured from the $0^{\rm th}$~moment maps; {\bf (6)}~Maximum circular velocity ($V_{\rm circ}=\sqrt{GM(<r)/r}$); {\bf (7)}~Circular velocity at $2\,{\rm kpc}$; {\bf (8)}~\HI\ gas azimuthal velocity at $2\,{\rm kpc}$; {\bf (9)}~Rotation velocity at $2\,{\rm kpc}$ as fit using \barolo.\label{tab_APprops}}
  \begin{tabular}{clllrrrrr}
    \hline
    \multicolumn{1}{c}{Symbol}&&\multicolumn{1}{c}{$M_{\rm HI}$}&\multicolumn{1}{c}{$M_\star$}&\multicolumn{1}{c}{$R_{\rm HI}$}&\multicolumn{1}{c}{${\rm max}(V_{\rm circ})$}&\multicolumn{1}{c}{$V_{\rm circ}(2\,{\rm kpc})$}&\multicolumn{1}{c}{$V_{\rm gas}(2\,{\rm kpc})$}&\multicolumn{1}{c}{$V_{\rm rot}(2\,{\rm kpc})$}\\
    \multicolumn{1}{c}{Number}&\multicolumn{1}{c}{Object}&\multicolumn{1}{c}{$({\rm M}_\odot)$}&\multicolumn{1}{c}{$({\rm M_\odot})$}&\multicolumn{1}{c}{$({\rm kpc})$}&\multicolumn{1}{c}{$({\rm km}\,{\rm s}^{-1})$}&\multicolumn{1}{c}{$({\rm km}\,{\rm s}^{-1})$}&\multicolumn{1}{c}{$({\rm km}\,{\rm s}^{-1})$}&\multicolumn{1}{c}{$({\rm km}\,{\rm s}^{-1})$}\\
    \hline
    $\mathbf{1}$&{\bf AP-L1-V1-4-0}&$2.6\times10^{9}$&$2.6\times10^{9}$&$23.8$&$91.0$&$68.2$&$60.1$&$51.4$\\
$\mathbf{2}$&{\bf AP-L1-V1-6-0}&$1.2\times10^{8}$&$8.8\times10^{8}$&$3.3$&$60.2$&$45.5$&$45.4$&$32.8$\\
$\mathbf{3}$&{\bf AP-L1-V1-7-0}&$3.0\times10^{8}$&$6.7\times10^{8}$&$4.2$&$72.0$&$61.6$&$59.4$&$41.0$\\
$\mathbf{4}$&{\bf AP-L1-V1-8-0}&$2.9\times10^{8}$&$7.7\times10^{8}$&$3.4$&$68.2$&$52.0$&$52.2$&$50.8$\\
$5$&AP-L1-V4-6-0&$4.4\times10^{8}$&$1.4\times10^{9}$&$7.5$&$86.4$&$74.8$&$62.6$&$25.2$\\
$\mathbf{6}$&{\bf AP-L1-V4-8-0}&$6.1\times10^{8}$&$8.1\times10^{8}$&$8.7$&$69.1$&$48.0$&$41.4$&$29.5$\\
$7$&AP-L1-V4-10-0&$2.2\times10^{9}$&$4.9\times10^{8}$&$17.5$&$66.2$&$45.5$&$25.3$&$25.0$\\
$8$&AP-L1-V4-13-0&$3.9\times10^{8}$&$2.7\times10^{8}$&$8.5$&$64.8$&$53.3$&$35.3$&$31.2$\\
$\mathbf{9}$&{\bf AP-L1-V4-14-0}&$1.0\times10^{9}$&$4.2\times10^{8}$&$11.8$&$60.2$&$44.7$&$40.5$&$31.4$\\
$10$&AP-L1-V6-5-0&$1.3\times10^{9}$&$2.3\times10^{9}$&$12.4$&$89.5$&$72.4$&$55.0$&$71.1$\\
$\mathbf{11}$&{\bf AP-L1-V6-6-0}&$2.1\times10^{9}$&$7.6\times10^{8}$&$10.3$&$68.0$&$46.4$&$39.7$&$31.1$\\
$\mathbf{12}$&{\bf AP-L1-V6-7-0}&$1.5\times10^{9}$&$5.1\times10^{8}$&$12.7$&$70.2$&$43.9$&$48.7$&$29.9$\\
$\mathbf{13}$&{\bf AP-L1-V6-8-0}&$2.8\times10^{9}$&$8.4\times10^{8}$&$21.2$&$75.9$&$55.6$&$48.0$&$43.3$\\
$14$&AP-L1-V6-11-0&$3.1\times10^{8}$&$7.5\times10^{8}$&$5.6$&$60.3$&$49.8$&$33.0$&$16.8$\\
$15$&AP-L1-V6-12-0&$1.7\times10^{9}$&$1.5\times10^{9}$&$16.3$&$76.6$&$61.8$&$51.3$&$38.8$\\
$16$&AP-L1-V6-15-0&$6.3\times10^{8}$&$3.4\times10^{8}$&$8.4$&$61.6$&$46.0$&$32.2$&$52.5$\\
$17$&AP-L1-V6-16-0&$4.1\times10^{8}$&$5.3\times10^{8}$&$7.9$&$65.5$&$55.0$&$44.1$&$58.8$\\
$\mathbf{18}$&{\bf AP-L1-V6-18-0}&$1.9\times10^{8}$&$3.0\times10^{8}$&$4.4$&$61.9$&$56.6$&$53.5$&$57.2$\\
$\mathbf{19}$&{\bf AP-L1-V6-19-0}&$6.5\times10^{8}$&$5.2\times10^{8}$&$9.2$&$61.0$&$46.7$&$50.7$&$26.0$\\
$\mathbf{20}$&{\bf AP-L1-V6-20-0}&$2.3\times10^{7}$&$3.8\times10^{8}$&$1.7$&$67.7$&$59.9$&$62.3$&$73.5$\\
$\mathbf{21}$&{\bf AP-L1-V10-5-0}&$2.0\times10^{9}$&$6.1\times10^{9}$&$12.9$&$108.8$&$83.3$&$80.5$&$68.6$\\
$22$&AP-L1-V10-6-0&$2.1\times10^{9}$&$4.0\times10^{9}$&$13.3$&$103.9$&$71.2$&$59.9$&$30.8$\\
$23$&AP-L1-V10-13-0&$6.0\times10^{8}$&$2.0\times10^{9}$&$5.6$&$83.6$&$74.9$&$59.6$&$50.1$\\
$24$&AP-L1-V10-14-0&$1.4\times10^{9}$&$9.7\times10^{8}$&$13.6$&$66.3$&$51.0$&$43.1$&$28.5$\\
$25$&AP-L1-V10-16-0&$1.1\times10^{9}$&$9.6\times10^{8}$&$12.5$&$75.5$&$51.6$&$35.5$&$43.2$\\
$26$&AP-L1-V10-17-0&$4.4\times10^{8}$&$7.6\times10^{8}$&$6.4$&$67.4$&$48.5$&$38.5$&$20.5$\\
$27$&AP-L1-V10-19-0&$4.9\times10^{8}$&$4.9\times10^{8}$&$9.3$&$67.2$&$48.2$&$26.9$&$20.9$\\
$28$&AP-L1-V10-20-0&$3.0\times10^{8}$&$7.0\times10^{8}$&$1.9$&$73.5$&$62.3$&$43.0$&$47.4$\\
$29$&AP-L1-V10-22-0&$3.9\times10^{8}$&$7.6\times10^{8}$&$7.2$&$65.5$&$48.0$&$31.2$&$22.4$\\
$\mathbf{30}$&{\bf AP-L1-V10-30-0}&$6.4\times10^{8}$&$3.6\times10^{8}$&$9.4$&$61.4$&$49.5$&$45.2$&$22.4$\\
$31$&AP-L1-V11-3-0&$4.6\times10^{9}$&$9.5\times10^{9}$&$29.1$&$118.4$&$94.6$&$79.5$&$81.3$\\
$\mathbf{32}$&{\bf AP-L1-V11-5-0}&$4.7\times10^{9}$&$3.2\times10^{9}$&$23.8$&$91.1$&$65.5$&$60.4$&$37.2$\\
$33$&AP-L1-V11-6-0&$4.1\times10^{9}$&$1.4\times10^{9}$&$28.6$&$88.5$&$67.6$&$47.3$&$43.2$\\

    \hline
  \end{tabular}
\end{table*}

\begin{table*}
  \caption{Selected properties of the \things\ and \littlethings\ galaxies. Columns: {\bf (1)}~Number corresponding to labels in Figs.~\ref{fig_size_mass}, \ref{fig_symmetry} \& \ref{fig_v2_vmax}; {\bf (2)}~Galaxy name used in survey publications; {\bf (3)}~Survey; {\bf (4)}~Distance; {\bf (5)}~Average inclination; {\bf (6)}~Average position angle; {\bf (7)}~Neutral hydrogen mass (the mass used in the BTFR is $M_{\rm bar} = M_\star + 1.4M_{\rm HI}$); {\bf (8)}~Stellar mass, assuming either `diet Salpeter' (\things) or Chabrier (\littlethings) IMF (galaxies with no reported measurement marked `--'); {\bf (9)}~\HI\ size defined as the radius where $\Sigma_{\rm HI}$ drops to $1\,{\rm M}_\odot\,{\rm pc}^{-2}$ (galaxies with surface density profiles that do not cross this value marked `--'); {\bf (10)}~Maximum rotation velocity; {\bf (11)}~Rotation velocity at $2\,{\rm kpc}$ (rotation curves with no measurements near $2\,{\rm kpc}$ marked `--'). {\bf References}~for all quantities are \citet{2008AJ....136.2563W,2011AJ....141..193O,2015AJ....149..180O}, except column (7) which is drawn from \citet{2008AJ....136.2563W,2012AJ....144..134H} and column (9) which we measure directly from the moment maps provided by the survey teams. \label{tab_thingsprops}}
  \begin{tabular}{clcrrrllrrr}
    \hline
    \multicolumn{1}{c}{Symbol}&&&\multicolumn{1}{c}{$D$}&\multicolumn{1}{c}{Incl.}&\multicolumn{1}{c}{PA}&\multicolumn{1}{c}{$M_{\rm HI}$}&\multicolumn{1}{c}{$M_\star$}&\multicolumn{1}{c}{$R_{\rm HI}$}&\multicolumn{1}{c}{${\rm max}(V_{\rm rot})$}&\multicolumn{1}{c}{$V_{\rm rot}(2\,{\rm kpc})$}\\
    \multicolumn{1}{c}{Number}&\multicolumn{1}{c}{Object}&\multicolumn{1}{c}{Survey}&$({\rm Mpc})$&\multicolumn{1}{c}{$(^\circ)$}&\multicolumn{1}{c}{$(^\circ)$}&\multicolumn{1}{c}{$({\rm M}_\odot)$}&\multicolumn{1}{c}{$({\rm M_\odot})$}&\multicolumn{1}{c}{$({\rm kpc})$}&\multicolumn{1}{c}{$({\rm km}\,{\rm s}^{-1})$}&\multicolumn{1}{c}{$({\rm km}\,{\rm s}^{-1})$}\\
    \hline
    $1$&CVnIdwA&LITTLE THINGS&$3.6$&$66$&$48$&$4.7\times10^{7}$&$4.9\times10^{6}$&$1.5$&$26.4$&$25.9$\\
$2$&DDO43&LITTLE THINGS&$7.8$&$41$&$294$&$1.7\times10^{8}$&\multicolumn{1}{c}{--}&$2.9$&$38.7$&$31.5$\\
$3$&DDO46&LITTLE THINGS&$6.1$&$28$&$274$&$1.9\times10^{8}$&\multicolumn{1}{c}{--}&--&$76.3$&$73.2$\\
$4$&DDO47&LITTLE THINGS&$5.2$&$46$&$312$&$3.9\times10^{8}$&\multicolumn{1}{c}{--}&$5.2$&$64.7$&$23.7$\\
$5$&DDO50&LITTLE THINGS&$3.4$&$50$&$176$&$7.1\times10^{8}$&$1.1\times10^{8}$&$0.1$&$38.8$&$31.2$\\
$6$&DDO52&LITTLE THINGS&$10.3$&$43$&$8$&$2.7\times10^{8}$&$5.4\times10^{7}$&$4.5$&$61.7$&$42.6$\\
$7$&DDO53&LITTLE THINGS&$3.6$&$27$&$132$&$5.2\times10^{7}$&$9.8\times10^{6}$&$1.4$&$32.0$&$29.3$\\
$8$&DDO70&LITTLE THINGS&$1.3$&$50$&$44$&$4.1\times10^{7}$&$1.9\times10^{7}$&$1.7$&$43.9$&$43.9$\\
$9$&DDO87&LITTLE THINGS&$7.7$&$56$&$235$&$2.5\times10^{8}$&$3.2\times10^{7}$&$2.7$&$56.6$&$28.0$\\
$10$&DDO101&LITTLE THINGS&$6.4$&$51$&$287$&$2.3\times10^{7}$&$6.6\times10^{7}$&$1.4$&$64.9$&$63.3$\\
$11$&DDO126&LITTLE THINGS&$4.9$&$65$&$138$&$1.4\times10^{8}$&$1.6\times10^{7}$&$3.1$&$38.7$&$30.7$\\
$12$&DDO133&LITTLE THINGS&$3.5$&$43$&$360$&$1.0\times10^{8}$&$3.0\times10^{7}$&$2.7$&$46.7$&$41.6$\\
$13$&DDO154&LITTLE THINGS&$3.7$&$68$&$226$&$2.9\times10^{8}$&$8.3\times10^{6}$&$3.7$&$51.1$&$35.8$\\
$14$&DDO168&LITTLE THINGS&$4.3$&$46$&$276$&$3.0\times10^{8}$&$5.9\times10^{7}$&--&$61.9$&$57.5$\\
$15$&DDO210&LITTLE THINGS&$0.9$&$67$&$65$&$2.0\times10^{6}$&$6.0\times10^{5}$&--&$12.0$&--\\
$16$&DDO216&LITTLE THINGS&$1.1$&$64$&$134$&$5.6\times10^{6}$&$1.5\times10^{7}$&$0.1$&$18.9$&$18.9$\\
$17$&F564-V3&LITTLE THINGS&$8.7$&$56$&$12$&$4.1\times10^{7}$&\multicolumn{1}{c}{--}&$0.9$&$39.2$&$38.7$\\
$18$&IC10&LITTLE THINGS&$0.7$&$47$&$56$&$6.0\times10^{7}$&\multicolumn{1}{c}{--}&--&$36.4$&--\\
$19$&IC1613&LITTLE THINGS&$0.7$&$48$&$74$&$3.4\times10^{7}$&$2.9\times10^{7}$&$0.0$&$21.1$&$20.5$\\
$20$&NGC1569&LITTLE THINGS&$3.4$&$69$&$122$&$2.5\times10^{8}$&$3.6\times10^{8}$&$2.9$&$39.3$&$36.6$\\
$21$&NGC2366&LITTLE THINGS&$3.4$&$63$&$39$&$6.9\times10^{8}$&$6.9\times10^{7}$&$5.6$&$59.8$&$41.9$\\
$22$&NGC3738&LITTLE THINGS&$4.9$&$23$&$292$&$1.1\times10^{8}$&$4.7\times10^{8}$&--&$132.7$&$125.6$\\
$23$&UGC8508&LITTLE THINGS&$2.6$&$82$&$126$&$1.9\times10^{7}$&$7.8\times10^{6}$&$0.7$&$46.1$&$46.1$\\
$24$&WLM&LITTLE THINGS&$1.0$&$74$&$174$&$7.1\times10^{7}$&$1.6\times10^{7}$&$1.8$&$38.5$&$35.1$\\
$25$&Haro29&LITTLE THINGS&$5.9$&$61$&$214$&$6.3\times10^{7}$&$1.4\times10^{7}$&$1.2$&$43.5$&$34.4$\\
$26$&Haro36&LITTLE THINGS&$9.3$&$70$&$248$&$1.4\times10^{8}$&\multicolumn{1}{c}{--}&$2.3$&$58.2$&$37.6$\\
$1$&NGC925&THINGS&$9.2$&$66$&$287$&$4.6\times10^{9}$&$1.0\times10^{10}$&--&$119.9$&$34.7$\\
$2$&NGC2403&THINGS&$3.2$&$63$&$124$&$2.6\times10^{9}$&$5.1\times10^{9}$&$11.8$&$143.9$&$97.4$\\
$3$&NGC2841&THINGS&$14.1$&$74$&$153$&$8.6\times10^{9}$&$1.3\times10^{11}$&$24.3$&$323.9$&--\\
$4$&NGC2903&THINGS&$8.9$&$65$&$204$&$4.4\times10^{9}$&$1.6\times10^{10}$&$16.9$&$215.5$&$120.1$\\
$5$&NGC2976&THINGS&$3.6$&$64$&$334$&$1.4\times10^{8}$&$1.8\times10^{9}$&$2.3$&$86.2$&$74.9$\\
$6$&NGC3031&THINGS&$3.6$&$59$&$330$&$3.6\times10^{9}$&$7.9\times10^{10}$&--&$259.8$&$242.2$\\
$7$&NGC3198&THINGS&$13.8$&$72$&$215$&$1.0\times10^{10}$&$2.5\times10^{10}$&$26.5$&$158.7$&$76.7$\\
$8$&IC2574&THINGS&$4.0$&$53$&$56$&$1.5\times10^{9}$&$1.0\times10^{9}$&$9.2$&$80.0$&$24.5$\\
$9$&NGC3521&THINGS&$10.7$&$73$&$340$&$8.0\times10^{9}$&$1.0\times10^{11}$&$19.9$&$233.4$&$192.1$\\
$10$&NGC3621&THINGS&$6.6$&$65$&$345$&$7.1\times10^{9}$&$1.6\times10^{10}$&$22.0$&$159.2$&$102.9$\\
$11$&NGC4736&THINGS&$4.7$&$41$&$296$&$4.0\times10^{8}$&$2.0\times10^{10}$&$3.4$&$198.3$&$168.7$\\
$12$&DDO154&THINGS&$4.3$&$66$&$230$&$3.6\times10^{8}$&$2.6\times10^{7}$&$4.5$&$50.0$&$34.6$\\
$13$&NGC5055&THINGS&$10.1$&$59$&$102$&$9.1\times10^{9}$&$1.3\times10^{11}$&$16.4$&$211.6$&$185.3$\\
$14$&NGC6946&THINGS&$5.9$&$33$&$243$&$4.2\times10^{9}$&$6.3\times10^{10}$&$15.9$&$224.3$&$132.5$\\
$15$&NGC7331&THINGS&$14.7$&$76$&$168$&$9.1\times10^{9}$&$1.6\times10^{11}$&$23.1$&$268.1$&$253.2$\\
$16$&NGC7793&THINGS&$3.9$&$50$&$290$&$8.9\times10^{8}$&$2.8\times10^{9}$&$6.7$&$117.9$&$76.2$\\
$17$&M81dwB&THINGS&$5.3$&$44$&$311$&$2.5\times10^{7}$&$3.0\times10^{7}$&$0.9$&$39.5$&$31.6$\\

    \hline
  \end{tabular}
\end{table*}

\section{Velocity field symmetry diagnostics}
\label{SecApp2}

\begin{figure*}
  {\leavevmode \includegraphics[width=2\columnwidth]{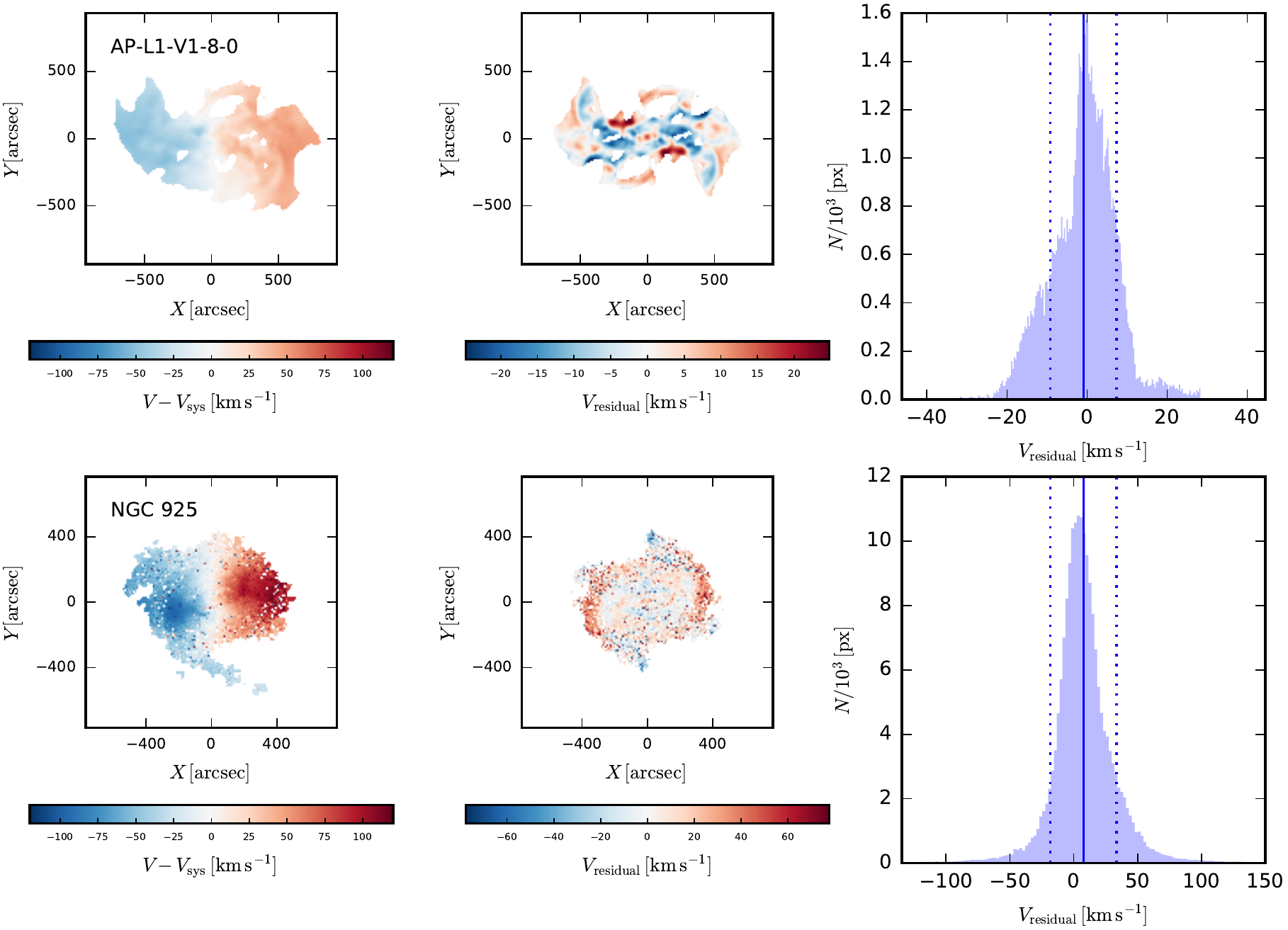}}
  \caption{Illustration of the measurements plotted in the centre panel of Fig.~\ref{fig_symmetry}. \emph{Left column: }$1^{\rm st}$~moment map for one simulated galaxy \mbox{AP-L1-V1-4-0} (above) and one \things\ galaxy NGC~4736 (below). \emph{Centre column: }The velocity field from the left column is rotated $180^\circ$ and aligned with the unrotated field by superimposing the galactic centre in each field. The two fields are then subtracted (with a sign change applied to the rotated field) to give the residual shown. \emph{Right column: }Histogram of the pixel values of the residual in the centre column. A perfectly rotationally symmetric velocity field would yield a sharp peak at $0\,{\rm km}\,{\rm s}^{-1}$. If one side of the galaxy has systematically higher $|V-V_{\rm sys}|$ than the other, the mean (vertical solid line) moves away from $0$; local asymmetries increase the rms width (vertical dotted lines) of the distribution.\label{fig_symmetry_ill}}
\end{figure*}

\begin{figure*}
  {\leavevmode \includegraphics[width=1.9\columnwidth]{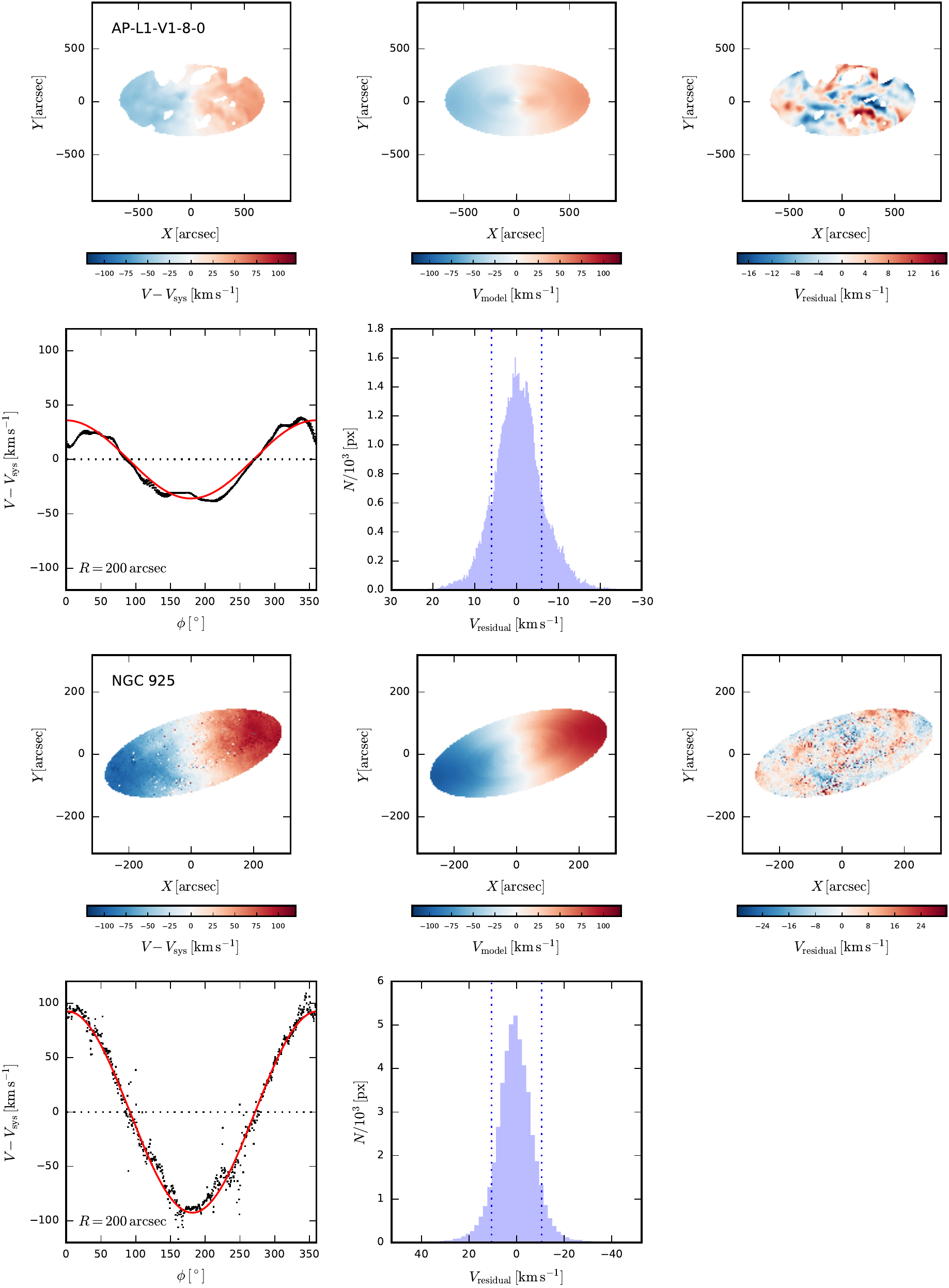}}
  \caption{Illustration of the measurement plotted in the right panel of Fig.~\ref{fig_symmetry}. There are two sets of five panels; the upper five correspond to the simulated galaxy \mbox{AP-L1-V1-4-0}, the lower five to the \things\ galaxy NGC~4736. \emph{Panel 1: }The velocity field of the galaxy, cropped to the radius enclosing $90$~per~cent of the \HI\ mass (simulated galaxy) or the maximum radius modelled in \citet[][or \citealp{2011AJ....141..193O,2015AJ....149..180O}, as appropriate]{2008AJ....136.2648D}. The orientation and aspect ratio of the ellipse are set by the global inclination and position angle of the galaxy. $(i,{\rm PA})=(60^\circ,270^\circ)$ for \apostle, for observed galaxies see Table~\ref{tab_thingsprops}. \emph{Panel 2: }A simple kinematic model constructed by fitting a cosine to the velocities in a series of rings (illustrated in panel 4). The inclinations and position angles are held fixed at the global values. \emph{Panel 3: }Residual after subtraction of the velocity fields in panels 1 \& 2. \emph{Panel 4: }Example of cosine fit to one ring at $R=300\,{\rm arcsec}$. The phase and amplitude are free parameters, but the vertical offset is fixed at $0\,{\rm km}\,{\rm s}^{-1}$. \emph{Panel 5: }Histogram of the pixel values of the residual in panel 3. Azimuthal asymmetries increase the rms width (vertical dotted lines) of the distribution; note that the width is the rms scatter from $0$.\label{fig_symmetry_ill_cosine}}
\end{figure*}

The centre and right panels of Fig.~\ref{fig_symmetry} show the results of measurements diagnosing the symmetry of the galaxy velocity fields which are most clearly explained graphically. In Fig.~\ref{fig_symmetry_ill} we illustrate the rotational symmetry diagnostic shown in the centre panel of Fig.~\ref{fig_symmetry}, and in Fig.~\ref{fig_symmetry_ill_cosine} we illustrate the azimuthal symmetry diagnostic shown in the right panel of Fig.~\ref{fig_symmetry}.

\section{Rotation curves of DDO~47 \& DDO~87}
\label{SecApp4}

\begin{figure*}
  {\leavevmode \includegraphics[width=1.5\columnwidth]{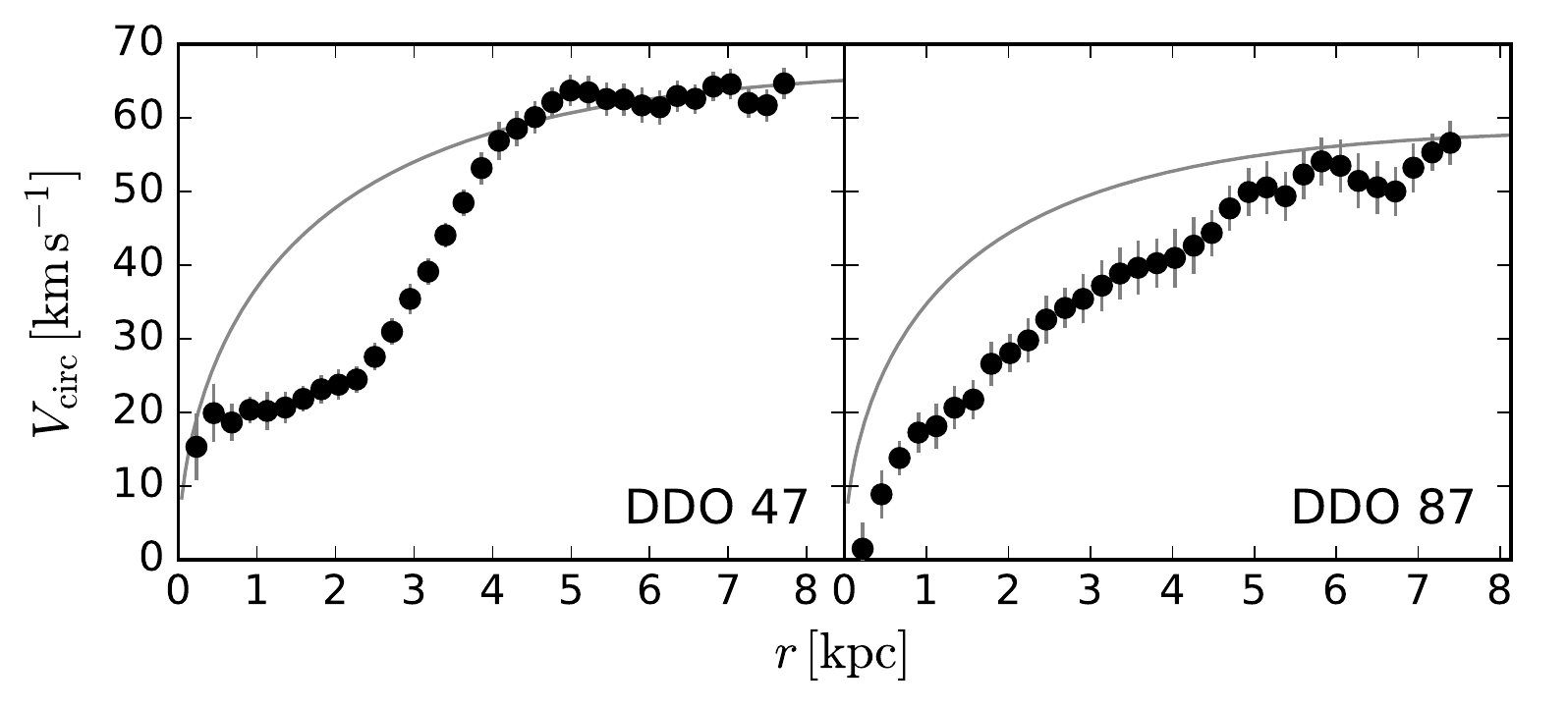}}
  \caption{Rotation curves of DDO~47 (left) and DDO~87 (right), as modelled by \citet{2015AJ....149..180O}. For comparison, we show the rotation curve of an NFW profile which goes through the last measured point, and which has parameters that lie on the mass--concentration relation of \citet{2014MNRAS.441..378L}. \label{fig_rotcurves}}
\end{figure*}

In Fig.~\ref{fig_rotcurves} we show the rotation curves of DDO~47 and DDO~87 \citep{2015AJ....149..180O}. Both are examples of slowly rising rotation curves which are often interpreted as evidence for dark matter `cores'. We illustrate the shape of a rotation curve corresponding to a `cuspy' profile with an NFW profile whose parameters lie on the mass--concentration relation of \citet{2014MNRAS.441..378L}.

\begin{figure}
  {\includegraphics[width=\columnwidth]{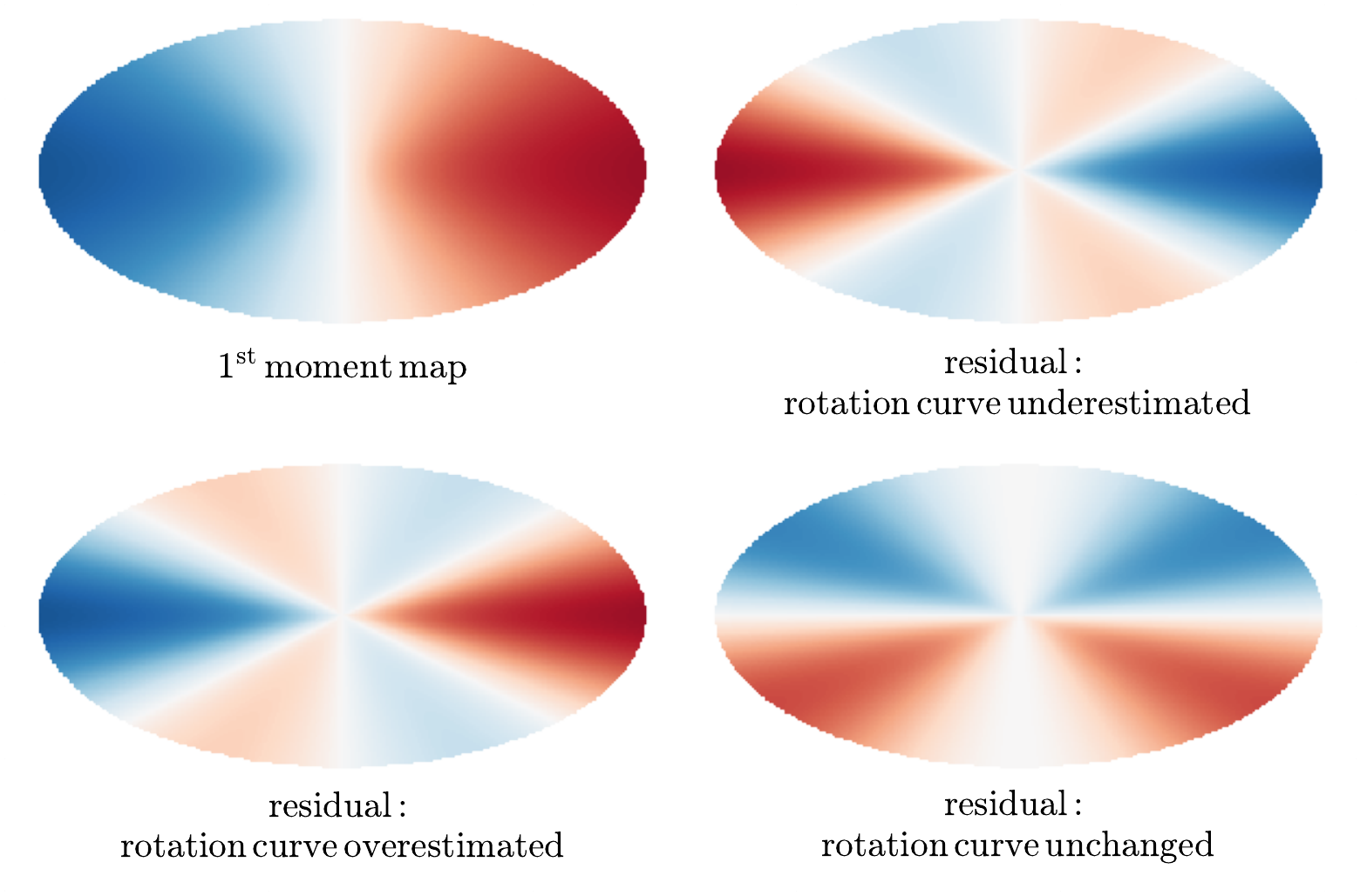}}
  \caption{Given a rotating disc oriented as illustrated in the upper left panel, an azimuthal $m=2$ modulation of the velocity field whose orientation is such that it causes an underestimate of the rotation curve (minima along the line of nodes) results in a three-peaked residual pattern, illustrated in the upper right panel. Note the residual peak (red in upper right) lying on the approaching side of the disc (blue in upper left). The case corresponding to an overestimate of the rotation curve (maxima along the line of nodes) is illustrated in the lower left panel; the residual is reflected relative to the case of an underestimate. In the case where the $m=2$ pattern causes no change to the rotation curve estimate (lower right), the residual is null along the line of nodes.\label{fig_schematic}}
\end{figure}

In Fig.~\ref{fig_schematic} we illustrate schematically the velocity field residuals corresponding to an azimuthal $m=2$ modulations causing an overestimate, underestimate, and no change to the rotation curve.

\section{\barolo\ configuration}
\label{SecApp3}

In Table~\ref{tab_baroloparams} we summarize the full configuration used for the \barolo\ software. We omit parameters which do not affect the result of the calculation (e.g. flags to enable or disable additional diagnostic output, file path definitions, etc.).

\begin{table*}
  \caption{Parameters used to configure the \barolo\ software, omitting parameters that have no impact on the result of the calculation (e.g., file paths, flags to enable optional outputs, etc.). Our parameter choices where they differ from the default values are inspired by the choices made by \citet{2017MNRAS.466.4159I}. These authors model the \littlethings\ galaxies; given that we mimic the observing setup of this survey and our simulated galaxies of interest are broadly similar to those in this sample, many of their parameter choices are applicable here. The most significant change we make is to restrict the inclination and position angles somewhat closer to their `true' values.\label{tab_baroloparams}}
  \begin{tabular}{lrlp{.5\linewidth}}
    \hline
    Parameter & Value & Units & Description \& comments \\
    \hline
    CHECKCHANNELS&  FALSE                & -- & Check for bad channels in the data cube?\\
    FLAGROBUSTSTATS&TRUE                 & -- & Use robust statistics?\\
    FLAGSEARCH &    FALSE                & -- & Search for sources in the data cube?\\
    FLAGRING  &     FALSE                & -- & Fit velocity field with a ring model?\\
    SMOOTH    &     FALSE                & -- & Smooth the data cube?\\
    GALFIT    &     TRUE                 & -- & Fit a 3D model to the data cube?\\
    BOX       &     NONE                 & ${\rm px}$ & Select a sub-region of the cube?\\
    NRADII    &     \emph{varies}        & -- & Number of rings to use; we use enough rings to reach the radius enclosing 90~per~cent of the \HI\ mass of the galaxy.\\
    RADSEP    &     $14.101$             & ${\rm arcsec}$ & Separation of rings.\\
    XPOS      &     \emph{varies}        & ${\rm px}$ & Centre of rings; set individually for each galaxy at the projected centre of the stellar light distribution.\\
    YPOS      &     \emph{varies}        & ${\rm px}$ & As XPOS.\\
    VSYS      &     $257.4528$           & ${\rm km}\,{\rm s}^{-1}$ & Systemic velocity.\\
    VROT      &     $30$                 & ${\rm km}\,{\rm s}^{-1}$ & Initial guess for rotation velocity.\\
    VDISP     &     $8$                  & ${\rm km}\,{\rm s}^{-1}$ & Initial guess for velocity dispersion.\\
    INC       &     $60$                 & ${\rm degrees}$ & Initial guess for inclination.\\
    DELTAINC  &     $15$                 & ${\rm degrees}$ & Allowed deviation of inclination from initial guess.\\
    PA        &     $270$                & ${\rm degrees}$ & Initial guess for position angle.\\
    DELTAPA   &     $20$                 & ${\rm degrees}$ & Allowed deviation of position angle from initial guess.\\
    Z0        &     $2.136$              & ${\rm arcsec}$ & Disc scale height.\\
    DENS      &     $-1$                 & ${\rm atoms}\,{\rm cm}^{-2}$ & Global column density of gas (unused when NORM LOCAL is set).\\
    FREE      &     VROT VDISP INC PA    & -- & Parameters to fit for each ring.\\
    MASK      &     SMOOTH               & -- & The data cube is smoothed by a factor of $2$ and a signal-to-noise cut is used to define a mask.\\
    BLANKCUT  &     $2.5$                & -- & Signal-to-noise threshold for mask construction.\\
    SIDE      &     B                    & -- & The entire galaxy is modelled; the approaching and receding sides can also be fit separately.\\
    NORM      &     LOCAL                & -- & The model is normalized pixel by pixel, i.e. the surface brightness is not explicitly fit.\\
    LTYPE     &     $1$                  & -- & Layer type along z is Gaussian.\\
    FTYPE     &     $2$                  & -- & Minimization function is $|{\rm model}-{\rm observed}|$.\\
    WFUNC     &     $1$                  & -- & Azimuthal weighting function is $|\cos\theta|$.\\
    TOL       &     $0.001$              & -- & Minimization tolerance.\\
    TWOSTAGE  &     TRUE                 & -- & Two stage fitting, i.e. geometric parameters are regularized and rotation velocity is fit again.\\
    POLYN     &     bezier               & -- & Degree of polynomial fitting INC and PA.\\
    FLAGERRORS&     \emph{varies}        & -- & Errors are estimated only for those galaxies we use for illustrative purposes.\\
    BWEIGHT   &     $1$                  & -- & Exponent of weight for blank pixels.\\
    LINEAR    &     $0.424$              & ${\rm channels}$ & Instrumental spectral broadening (standard deviation).\\
    CDENS     &     $10$                 & -- & Number of clouds to use in building the ring model.\\
    NV        &     $200$                & -- & Number of sub-clouds used within each cloud (see CDENS) to populate the spectral axis of the ring model.\\
    \hline
  \end{tabular}
\end{table*}

\section{Supplementary materials}
\label{AppSupp}
We include as Supplementary Materials, available online, the first three moment maps (similar to Fig.~\ref{fig_mom_maps}), tilted ring modelling summary (similar to Fig.~\ref{fig_barolo_fits}), and position-velocity diagrams for the $33$~\apostle\ galaxies in the sample defined in Sec.~\ref{SubsecSample}.

\label{lastpage}

\end{document}